\documentclass[reprint,showkeys, amsmath,amssymb, aps, floatfix]{revtex4-2}

\usepackage{graphicx,subfigure}
\usepackage{dcolumn}
\usepackage[outdir=./epsTopdf/]{epstopdf}
\usepackage{bm}
\usepackage{amsmath}
\usepackage{amssymb}
\usepackage{physics}
\usepackage{xcolor}
\usepackage{slashed}
\usepackage{multirow}
\usepackage{float}
\usepackage{capt-of}
\usepackage{comment}
\usepackage{ mathrsfs }
\usepackage[colorlinks=true,linkcolor=blue,urlcolor=blue,citecolor=blue]{hyperref}

\DeclareUnicodeCharacter{2212}{-}
\begin{document}
	
	\preprint{APS/123-QED}
	
	\title{Sensitivity of polarizations and spin correlations of $Z$ boson to anomalous \\neutral triple gauge couplings at lepton collider with polarized beams}
	
\author{Amir Subba}
\email{as19rs008@iiserkol.ac.in}

\author{Ritesh K. Singh}
\email{ritesh.singh@iiserkol.ac.in}
\affiliation{Department of Physical Sciences, Indian Institute of Science Education and Research Kolkata, Mohanpur, 741246, India\\
	}
	
\date{\today}

 \begin{abstract}
    We investigate the effects of anomalous neutral triple gauge couplings in $ZZ$ and $Z\gamma$ production processes, followed by the leptonic decay of the $Z$ boson, at a lepton collider with center-of-mass energy $\sqrt{s}=250$ GeV and polarized beams. We use an effective Lagrangian formalism to parameterize the anomalous couplings in terms of dimension-8 operators $c_{\widetilde{B}W}$, $c_{BW}$, $c_{WW}$, and $c_{BB}$, and study the sensitivity of observables such as cross~section, polarization, and spin correlation as functions of these couplings. We perform a Bayesian statistical analysis using Markov Chain Monte Carlo methods to determine simultaneous limits on the anomalous couplings, taking into account various luminosities $\mathcal{L} \in \{0.1~\text{ab}^{-1}, 0.3~\text{ab}^{-1}, 1~\text{ab}^{-1}, 3~\text{ab}^{-1}, 10~\text{ab}^{-1}\}$ and systematic uncertainties. We find that polarization and spin correlation observables significantly enhance the sensitivity to anomalous couplings, providing stringent constraints on these couplings.
    \end{abstract}
\maketitle
\section{Introduction}
\label{sec:intro}
The Standard Model~(SM) of particle physics is a marvel of scientific achievement. It has undergone rigorous experimental scrutiny to emerge as the most extensively tested theory of fundamental particles and their interactions. The discovery of scalar $J^P=0^+$ boson by ATLAS~\cite{ATLAS:2012yve} and CMS~\cite{CMS:2012qbp} collaborations at LHC consistent with the Standard Model Higgs boson completes the particle spectrum of SM. However, some unresolved issues exist within the SM framework, including fine-tuning the Higgs boson mass, which is susceptible to higher-order quantum corrections that could shift the mass away from the experimental value of 125 GeV. Another unresolved issue is the strong-CP problem, which refers to the unexplained value of the theta parameter ($\theta$) in the quantum chromodynamics (QCD) sector of the SM.
Furthermore, dark matter~\cite{Planck:2013pxb}, which constitutes approximately $85\%$ of the matter in the Universe, remains a mystery, and its structure still needs to be fully understood. Explaining the non-zero mass of neutrinos and the transformation between the three generations of neutrinos requires physics beyond SM. Recent measurements of the $W$ boson mass~\cite{CDF:2022hxs} and the magnetic moment of the muon~\cite{Muong-2:2021ojo} have shown significant deviations from the SM. Several models have been proposed to address these shortcomings, such as supersymmetry, technicolor, universal extra dimensions, and string theory, which incorporate extra dimensions, new symmetries, and new particles. However, no experimental evidence has yet confirmed any of these models.
\\\\ 
In the absence of any signature in favor of specific models, one adopts a model-independent framework to parameterize the effects beyond the Standard Model, known as the effective field theory~(EFT)~\cite{Georgi:1993mps,Weinberg:1978kz,Weinberg:1980wa} approach. This approach extends the SM Lagrangian by adding higher-order gauge invariant terms constructed from SM states. In general, if a field theory contains a set of fields $A$ and $B$, where $A$ is light, and $B$ are heavy states, then the action $\widetilde{\mathcal{S}}\left[A\right]$ of the effective field theory may be obtained from the action $\mathcal{S}\left[A,B\right]$ of the full field theory by a functional integral over $B$,
$e^{\left(i\widetilde{\mathcal{S}}\left[A\right]\right)} = \int \left[\text{d}B\right]e^{i\mathcal{S}\left[A,B\right]}$. One of the issues with this indirect formalism is the possibility of matching low energy theory to multiple UV complete theory. Such ambiguities are removed only with the discovery of new particles. Usually, the EFT approach remains valid upto some characteristic energy scale, $\Lambda$, such that $\Lambda > M_{\text{EW}}$, $M_{\text{EW}}$ is the electroweak scale. Each higher dimensional terms~($d > 4$) are weighted by $\Lambda^{d-4}$ power, such that the effective Lagrangian is,
\begin{equation}
\label{eqn:eftLag}
    \mathscr{L}_{\text{eff}} = \mathscr{L}_{\text{SM}} + \frac{1}{\Lambda^2}\sum_ic_i^{(6)}\mathscr{O}_i^{(6)} + \frac{1}{\Lambda^4}\sum_ic_i^{(8)}\mathscr{O}_i^{(8)} + ..,
\end{equation}
where the sum over $i$ runs over a basis $\{\mathscr{O}_i^{(d)}\}$ in the $d\in\{6,8,10,..\}$-dimension gauge-invariant operator space and  $c^\prime s$ are Wilson coefficients related to those operators. Since the scale of new physics is in several TeV, the higher order terms are highly suppressed; thus, we can safely truncate the expansion in Eq.~(\ref{eqn:eftLag}) to some lowest order for collider energy of few TeVs. The presence of these operators affects various structures like couplings of bosons and fermions, and the vacuum expectation value of the Higgs field. The deviation in these structures can be experimentally probed through measurements of various observables.  \\\\
The SM non-abelian $SU(2)\times U(1)$ gauge structure induces triple and quartic gauge boson couplings. Studies of triple and quartic couplings between the gauge bosons test the SM description of gauge sector interactions and provide sensitivity to physics beyond the SM by examining production rates and kinematics. In the case of triple gauge couplings, the structure SM allows is $W^-W^+V,~V\in\{\gamma, Z\}$, and no neutral triple gauge couplings~(NTGC) at tree level. At the lowest order, such charged triple gauge boson couplings are affected by dimension-6 operators, while NTGC is generated by dimension-8 operators~\cite{Degrande:2013kka} at the tree level. Besides operator formalism, anomalous NTGC~(aNTGC) can also be parameterized as effective vertex. Assuming only Lorentz and $U(1)_{\text{em}}$ gauge invariance, the most general $V_1V_2V_3$ vertex function, where $V_{2,3}$ are on-shell neutral gauge bosons, while $V_1\in\{Z^\star,\gamma^\star\}$ is off-shell, can be written as~\cite{Hagiwara:1986vm,Gounaris:1999kf},
\begin{equation}
    \begin{aligned}
        ie\Gamma^{\alpha\beta\gamma}_{VZZ}(q_1,q_2,q_3) &= \frac{-e(q_1^2-m_V^2)}{m_Z^2}\left[f_4^V\left(q_1^\alpha g^{\mu\beta}+q_1\beta g^{\mu\alpha}\right)\right.\\&-\left. f_5^V\epsilon^{\mu\alpha\beta\rho}\left(q_2-q_3\right)_\rho\right],\\
        ie\Gamma^{\alpha\beta\mu}_{VZ\gamma}(q_1,q_2,q_3)&= \frac{-e(q_1^2-m_V^2)}{m_Z^2}\left[h_1^V\left(q_3^V g^{\alpha\beta}-q_3^\alpha g^{\mu\beta}\right) \right.\\&+\left. \frac{h_2^V}{m_Z^2}q_1^\alpha\left(q_1q_3 g^{\mu\beta}-q^\mu_3 q^\beta_1\right)\right.\\&-\left. h_3^V\epsilon^{\mu\alpha\beta\rho}q_{3\rho}-\frac{h_4^V}{m_Z^2}q_1^V\epsilon^{\mu\beta\rho\sigma}q_{1\rho}q_{3\sigma}\right], 
    \end{aligned}
\end{equation}
where $V \in \{Z,\gamma\}$. The effective Lagrangian generating the above two vertices is,
\begin{widetext}
\begin{equation}
\label{eqn:Lageff}
    \begin{aligned}
        \mathscr{L} &= \frac{e}{m_Z^2}\left[-[f_4^\gamma (\partial_\mu F^{\mu\beta}) + f_4^Z(\partial_\mu Z^{\mu\beta})]Z_\alpha(\partial^\alpha Z_\beta) + [f_5^\gamma(\partial^\sigma F_{\sigma\mu}) + f_5^Z(\partial^\sigma Z_{\sigma\mu})]\widetilde{Z}^{\mu\beta}Z_\beta - [h_1^\gamma(\partial^\sigma F_{\sigma\mu}) + h_1^Z(\partial^\sigma Z_{\sigma\mu})]Z_\beta F^{\mu\beta} \right.\\&-\left. [h_3^\gamma(\partial_\sigma F^{\sigma\rho}) + h_3^Z(\partial_\sigma Z^{\sigma\rho})]Z^\alpha \widetilde{F}_{\rho\alpha} - \left\{\frac{h_2^\gamma}{m_Z^2}[\partial_\alpha\partial_\beta\partial^\rho F_{\rho\mu}]+\frac{h_2^Z}{m_Z^2}[\partial_\alpha\partial_\beta(\Box +m_Z^2)Z_\mu]\right\}Z^\alpha F^{\mu\beta} \right.\\&+\left. \left\{\frac{h_4^\gamma}{2m_Z^2}[\Box\partial^\sigma F^{\rho\alpha}]+\frac{h_4^Z}{2m_Z^2}[(\Box + m_Z^2)\partial^\sigma Z^{\rho\alpha}]\right\}Z_\sigma\widetilde{F}_{\rho\alpha}\right],
    \end{aligned}
\end{equation}
\end{widetext}
where $\widetilde{V}_{\mu\nu} = \frac{1}{2}\epsilon_{\mu\nu\rho\sigma}V^{\rho\sigma}\left(\epsilon^{0123}=+1\right)$ and the field tensor is defined as $Z_{\mu\nu} = (\partial_\mu Z_\nu - \partial_\nu Z_\mu)$ and $F_{\mu\nu} = (\partial_\mu A_\nu - \partial_\nu A_\mu)$. The couplings $f_4^V,h_1^V,h_2^V$ correspond to the $CP$-odd tensorial structures, while $f_5^V,h_3^V,h_1^V$ corresponds to the $CP$-even ones. The above vertex formalism is used to parameterize aNTGC by many experiments like LEP~\cite{L3:1999tgr,OPAL:2003gfi,ALEPH:2006bhb}, LHC~\cite{CMS:2012exm,ATLAS:2012bpb,atlascollaboration2023}, and Tevatron~\cite{D0:2007awp,CDF:2011ab}.\\
Following Ref.~\cite{Degrande:2013kka}, there are one $CP$-even and three $CP$-odd dimension-8 operators that generates anomalous $ZVV,V\in\{Z^\star,\gamma^\star\}$ couplings, they are,
\begin{equation}
\label{eqn:dim8op}
    \begin{aligned}
        \mathscr{O}_{\widetilde{B}W} &= i\Phi^\dagger\widetilde{B}_{\mu\nu}W^{\mu\rho}\{D_\rho,D^\nu\}\Phi, \\
        \mathscr{O}_{BW} &= i\Phi^\dagger B_{\mu\nu}W^{\mu\rho}\{D_\rho,D^\nu\}\Phi, \\
        \mathscr{O}_{WW} &= i\Phi^\dagger W_{\mu\nu}W^{\mu\rho}\{D_\rho,D^\nu\}\Phi, \\
        \mathscr{O}_{BB} &= i\Phi^\dagger B_{\mu\nu}B^{\mu\rho}\{D_\rho,D^\nu\}\Phi.
    \end{aligned}
\end{equation}
Here, $\Phi$ is the Higgs doublet, covariant derivative, $D_\mu~=~\partial_\mu + \frac{i}{2}g^\prime B_\mu + ig \frac{\sigma^a}{2}W^a_\mu$, field tensor is defined as, $B_{\mu\nu}=\partial_\mu B_\nu - \partial_\nu B_\mu$ and $W_{\mu\nu}~=~\left(\partial_\mu W^a_\nu - \partial_\nu W^a_\mu + g\epsilon_{abc}W^b_\mu W^c_\nu\right)$. In Eq.~(\ref{eqn:dim8op}), the first operator is $CP$-even, and the rest are all $CP$-odd operators. These operators are suppressed by a fourth power of the new physics scale $\Lambda$, which is assumed to be in several TeV, implying the deviation due to these operators are tiny. The matrix element in the presence of these operators is
\begin{equation}
    |M|^2 = |M_{SM}|^2 + 2\mathscr{R}\left(M_{SM}M^\star_{8}\right) + |M_8|^2.
\end{equation}
Most of the contribution to the anomalous factor comes from the interference of SM with dimension-8 operators, which is the second term of the above equation. In this article, we work to the order of $1/\Lambda^{8}$ contribution from dimension-8 operators while keeping all other higher-order operator parameters to zero. Despite not inducing aNTGC at the tree level, the dimension-6 operators can nevertheless have an impact at the one-loop level. The one-loop contributions from the dimension-6 operators would be of the order of $(\alpha/4\pi)(s/\Lambda^2)$, while the contribution from the dimension-eight operators at the tree level would be of the order of $(sv^2/\Lambda^4)$. Due to this, for $\Lambda \leq 2v\sqrt{\pi/\alpha} \approx 10$~TeV, the contribution of the dimension-eight operators outweighs the one-loop contribution of the dimension-6 operators~\cite{Degrande:2013kka} and hence ignored in this work.\\\\
The anomalous parameters $f^\prime s$ and $h^\prime s$ of the Eq.~(\ref{eqn:Lageff}) can be expressed using the Wilson coefficient of dimension-8 operator. We define the related Wilson coefficient of effective operators in Eq.~(\ref{eqn:dim8op}) as,
\begin{equation}
c_i\in\{c_{\widetilde{B}W},c_{BW},c_{WW},c_{BB}\}.
\end{equation}
For a process with two on-shell $Z$ bosons, the $CP$-even couplings are related to Wilson coefficient $c_i$ of effective operators as~\cite{Degrande:2013kka},
\begin{equation}
\label{eqn:EFTLag1}
    \begin{aligned}
        &f_5^Z = 0,\\
        &f_5^\gamma = \frac{v^2m_Z^2}{4c_Ws_W}\frac{c_{\widetilde{B}W}}{\Lambda^4},
    \end{aligned}
\end{equation}
where $c_W$ and $s_W$ are cosine and sine of weak mixing angle, respectively. The $CP$-odd couplings are translated via,
 \begin{equation}   
    \label{eqn:EFTLag2}
    \begin{aligned}
       &f_4^Z = \frac{m_Z^2 v^2}{2c_Ws_W}\left[c_W^2\frac{c_{BB}}{\Lambda^4}+2c_Ws_W\frac{c_{BW}}{\Lambda^4}+4s_W^2\frac{c_{WW}}{\Lambda^4}\right], \\
        &f_4^\gamma = -m_Z^2v^2\left[-c_Ws_W\frac{c_{BB}}{\Lambda^4}+\frac{c_{BW}}{\Lambda^4}\left(c_W^2-s_W^2\right)\right.\\&+\left.4c_Ws_W\frac{c_{WW}}{\Lambda^4}\right].
    \end{aligned}
\end{equation}
For one on-shell $Z$ boson and one on-shell $\gamma$, the $CP$-even couplings are
 \begin{equation}   
    \label{eqn:EFTLag3}
    \begin{aligned}
        &h_3^Z = \frac{v^2m_Z^2}{4c_Ws_W}\frac{c_{\widetilde{B}W}}{\Lambda^4} \\
        &h_4^Z=h_3^\gamma=h_4^\gamma=0,
   \end{aligned}
\end{equation}
while the $CP$-odd couplings are related as,
 \begin{equation}   
    \label{eqn:EFTLag4}
    \begin{aligned}
        & h_1^Z = \frac{m_Z^2v^2}{4c_Ws_W}\left[-c_Ws_W\frac{c_{BB}}{\Lambda^4}+\frac{c_{BW}}{\Lambda^4}\left(c_W^2-s_w^2\right)\right.\\&+\left.4c_Ws_W\frac{c_{WW}}{\Lambda^4}\right],\\
        &h_1^\gamma = \frac{-m_Z^2v^2}{4c_W s_W}\left[s_W^2\frac{c_{BB}}{\Lambda^4}-2s_W c_W\frac{c_{BW}}{\Lambda^4}+4c_W^2\frac{c_{WW}}{\Lambda^4}\right],\\
        &h_2^Z = h_2^\gamma = 0.\\
    \end{aligned}
\end{equation}
It is useful to notice that not all anomalous couplings, $f_i^V,h_i^V$, are independent; given two processes, there are specific relations between those couplings that hold. This is evident from Eq.~(\ref{eqn:EFTLag1})-(\ref{eqn:EFTLag4}).
\begin{equation}
    \label{eqn:anomrel}
    \begin{aligned}
        f_5^\gamma &= h_3^Z,\\
        f_4^\gamma &= 4c_Ws_W h_1^Z.        
    \end{aligned}
\end{equation}
Thus the four Wilson's coefficient of dimension-8 operators can be related to and translated to four independent anomalous couplings,$\{f_5^\gamma, f_4^\gamma, f_4^Z, h_1^\gamma\}$. We discuss the behaviour of these anomalous couplings in Appendix~\ref{sec:app1}. The effect of higher order operators of Eq.~(\ref{eqn:dim8op}) is studied by creating a Universal FeynRules Output~(UFO)~\cite{Degrande:2011ua} model files by implementing those operators in FeynRules~\cite{Alloul:2013bka}. This model file was used for event generation in {\tt MadGraph5$\_$aMC$@$NLO}~\cite{Alwall:2011uj,Alwall:2014hca} in the presence of aNTGC. The input parameters of the model are:
\begin{equation}
    \begin{aligned}
        &m_t = 172.0~\text{GeV}, \quad m_Z = 91.187~\text{GeV},\\ & m_H = 125.0~\text{GeV}, \quad
        \alpha_{\text{EW}} = \frac{1}{127.9},\\ &G_F = 1.166\times 10^{-5}~\text{GeV}^{-2}, \alpha_s = 0.118.
    \end{aligned}
\end{equation}
This article aims to provide the sensitivity of various observables like cross~section, polarization, and spin correlation asymmetries to the anomalous couplings $c_i$ and to obtain the bounds on these couplings. Similar studies~\cite{Subba:2022czw,Rahaman:2016pqj,Rahaman:2017qql} used these spin-related observables to constrain anomalous couplings in the neutral and charged sector. The studies had shown significant sensitivity of such observables to new couplings in the case of charged triple gauge couplings~\cite{Subba:2022czw}. \\\\
Future lepton colliders like ILC~\cite{ILC:2013jhg,Adolphsen:2013kya}, CLIC~\cite{CLICdp:2018cto}, and FCC-ee~\cite{FCC:2018evy} will collide the polarized beams, increasing the signal statistics by improving the signal vs. background ratio. Due to the differing weak quantum numbers of left- and right-chiral fermions, they couple differently with the weak gauge bosons. In general, the polarized cross~section is written as:
\begin{equation}
    \sigma(\eta_3,\chi_3) = \frac{(1+\eta_3)(1+\chi_3)}{4}\sigma_{\text{LR}} + \frac{(1-\eta_3)(1-\chi_3)}{4}\sigma_{\text{RL}},
\end{equation}
where $\eta_3$ and $\chi_3$ are the longitudinal degree of polarization of electron and positron beam, respectively, and $\sigma_{\text{LR}}(\sigma_{\text{RL}})$ are cross~section for 100$\%$ left polarized~(right polarized) electron beam and 100$\%$ right polarized~(left polarized ) positron beam. The effect of initial beam polarization for $ZZ$ production at $e^-e^+$ collisions are also studied at one loop electroweak corrections together with soft and hard QED radiation~\cite{Demirci:2022lmr}. \\\\ 
Here, we probe $ZZ$, and $Z\gamma$ production followed by the leptonic decay of $Z$ bosons in a $e^-e^+$ collider, the process is defined as:
\begin{equation}
\label{process}
    e^- + e^+ \to Z + Z \to e^-e^+\mu^-\mu^+,
\end{equation}
and
\begin{equation}
    e^- + e^+ \to Z + \gamma \to l^-l^+\gamma.
\end{equation}
\begin{figure}[!h]
    \centering
    \includegraphics[width=0.23\textwidth]{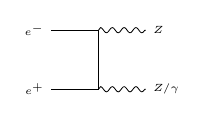}
    \includegraphics[width=0.23\textwidth]{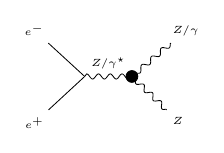}
    \caption{Representative Feynman diagrams for the production of two neutral boson in SM~(left panel) and presence of new physics would also induce a triple neutral gauge boson coupling shown in black blob in right panel. }
    \label{fig:feynman}
\end{figure}\\
The degree of polarization used are $\pm 80\%$ for electron and $\pm 30\%$ for positron. The values are taken to be commensurate with the initial plan of ILC~\cite{ILC:2013jhg}. Though we limit the current study to longitudinal polarized beams, the literature on transverse polarized beams can be found in Refs.~\cite{Swartz:1987xme,Chen:2022rgo,Hikasa:1984ng,Hikasa:1985qi,Renard:1988qg,Wen:2023xxc,Placidi:1985it}. 
The $ZZ\to 4l$ process in SM happens through electron exchange $t$ and $u$ channel, shown in left panel of Fig.~\ref{fig:feynman}. Additional, $\gamma/Z$ $s$-channel diagrams are included in presence of dimension-8 operators defined in Eq.~(\ref{eqn:dim8op}), which generates neutral triple gauge couplings~($ZZZ^\star,ZZ\gamma^\star$), shown in right panel of Fig.~\ref{fig:feynman}. The $ZZ\to 4l$ process is minimally diluted by background, and the events are kinematically reconstructable but suffer low statistics due to small branching fractions. As for $l^-l^+\gamma,~l^- \in \{e^-,\mu^-\}$, it can happen with the photon as a result of initial state radiation~(ISR) and leptons from the decay of $s$-channel mediated by $Z$ boson and $\gamma$, apart from the signal in consideration which is the production of $Z\gamma$ through $s$-channel mediated by $Z$ and $\gamma$. Some other diagrams contributing to these kinds of final state topology at the leading order are the production of pair of lepton through $s$-channel and subsequent final state radiation~(FSR). Also, it can proceed with the pair production of photons, with one of the photons being highly off-shell. However, we can suppress non-resonance contributions by imposing the $Z$-pole event selection condition on $e^-e^+$ invariant mass. We do not include the detector effects in our analysis for simplicity and because such effects are relatively smaller for $e^\pm, \mu^\pm$, and photon.
\\ \\
The production of $ZZ$ diboson leading to $4l$ final state has been studied at Large Electron-Positron~(LEP) Collider~\cite{ALEPH:1999ytd,ALEPH:2006bhb,ALEPH:2013dgf,DELPHI:2003kgg,L3:1999tgr,OPAL:2003gfi} where limits on aNTGC are provided in terms of parameters given in Eq.~(\ref{eqn:Lageff}). Several studies were also presented CDF~\cite{CDF:2011ab,CDF:2014nef}, D0~\cite{D0:2007awp,D0:2013rca}, ATLAS~\cite{ATLAS:2017bcd,ATLAS:2018nci} and CMS~\cite{CMS:2017dzg} collaborations. On the phenomenological side, aNTGC is extensively studied in Refs.~\cite{Rahaman:2016pqj,Rahaman:2017qql,Yilmaz:2019cue,Degrande:2013kka,Senol:2018cks,Spor:2022hhn,Jahedi:2022duc,Atag:2004cn,Ots:2004hk,Gutierrez-Rodriguez:2008rvy,Ellis:2019zex,Senol:2019qyl,Ellis:2022zdw,Hernandez-Juarez:2021mhi,Hernandez-Juarez:2022kjx,Rahaman:2018ujg}.
The list of tightest one parameter limits at $95\%$ confidence level~(CL) obtained at LHC by ATLAS collaboration in $ZZ \to 4l$ and $Z\gamma \to \nu\bar{\nu}\gamma$ final events is given in Table~\ref{tab:const}. 
\begin{table}[!h]
    \centering
    \caption{ \label{tab:const}Observed one dimensional $95\%$ confidence level~(CL) limits on $c_{\widetilde{B}W}$, $c_{BW}$, $c_{WW}$ and $c_{BB}$ EFT parameters from ATLAS collaboration in LHC.}
   \begin{tabular}{lll}\\ \hline \hline
    \multirow{2}{*}{Parameters~($c_i$)}&\multicolumn{2}{c}{Limits~(TeV$^{-2}$)}\\
    \cline{2-3}
    &$ZZ\to 4l$~\cite{ATLAS:2017bcd}&\quad$Z\gamma\to \nu\bar{\nu}\gamma $~\cite{ATLAS:2018nci}\\   \hline
      $c_{\widetilde{B}W}/\Lambda^4$&$ \left[-5.9,+5.9\right]$&\quad$ \left[-1.10,+1.10\right]$ \\  
      $c_{BW}/\Lambda^4$&$\left[-3.0,+3.0\right]$&\quad$ \left[-2.30,+2.30\right]$ \\
       $c_{WW}/\Lambda^4$&$\left[-3.3,+3.3\right]$&\quad$ \left[-0.64,+0.64\right]$ \\
       $c_{BB}/\Lambda^4$&$\left[-2.7,+2.8\right]$&\quad$ \left[-0.24,+0.24\right]$ \\\hline
       \hline
    \end{tabular}
\end{table}\\
The aim of the present work is to study the potential of future lepton collider i.e, ILC to probe the various higher dimensional operators inducing aNTGC discussed above at $\sqrt{s} = 250$~GeV. We exploit the spin related observables like polarizations and spin correlations to constrain the anomalous parameters. The construction of those observables in $\nu\bar{\nu}\gamma$ events becomes non-trivial due to the combinatorial issue in reconstructing two missing neutrinos. In the case of $ZZ\to4l$, we reconstruct the polarizations of the $Z$ bosons as well as the spin correlations of two $Z$ bosons. While for the $Z\gamma$ process, we focus only on the polarization of the $Z$ boson as the reconstruction of polarizations of a photon is not possible in collider experiments. \\\\
The plan of the paper is as follows: in Sec.~\ref{sec:spinobs}, we discuss the spin of a particle and the observables obtained using the spin. We focus on the asymmetries related to the polarization and spin-spin correlation of $Z$ bosons. In Sec.~\ref{sec:anomalous}, we discuss the limits of anomalous couplings. We conclude in Sec.~\ref{sec:conclude}.

\section{Observables: polarizations and spin-spin correlations}
\label{sec:spinobs}
 \begin{figure*}[!tbh]
     \centering
     \includegraphics[width=0.325\textwidth]{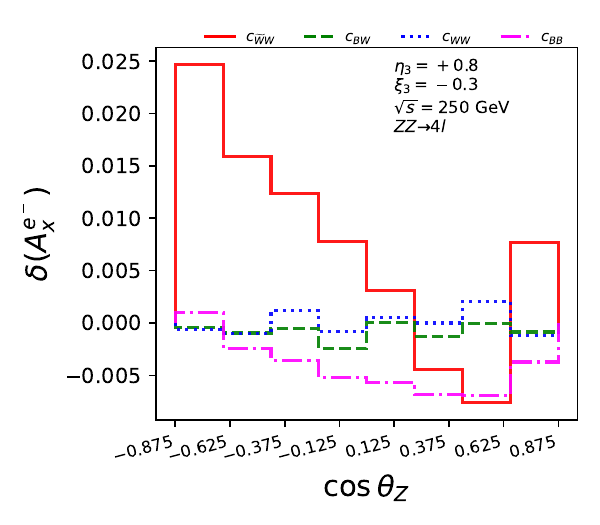}
     \includegraphics[width=0.325\textwidth]{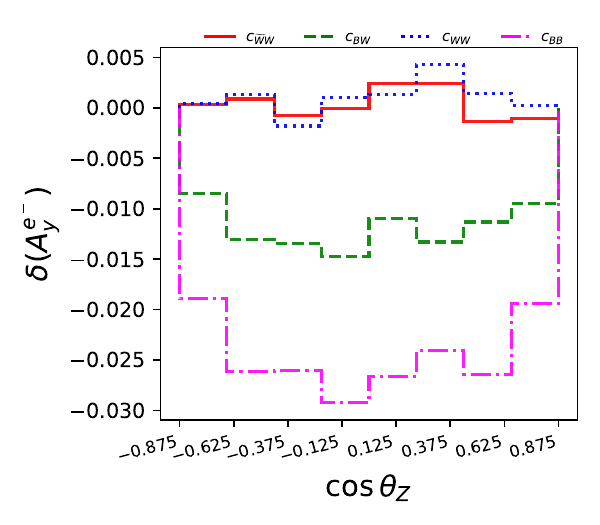}
     \includegraphics[width=0.325\textwidth]{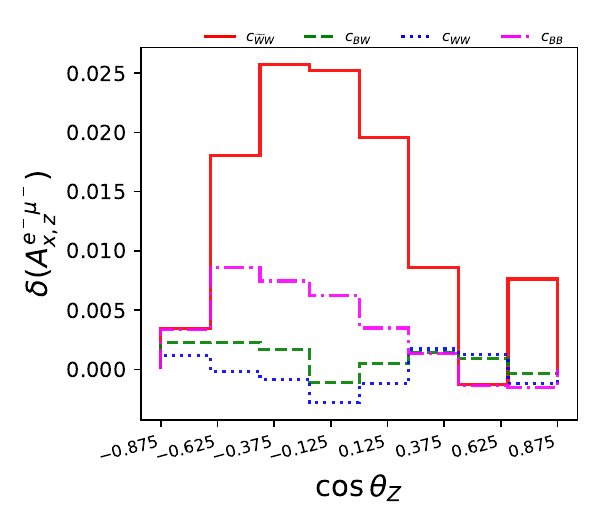}
     \includegraphics[width=0.325\textwidth]{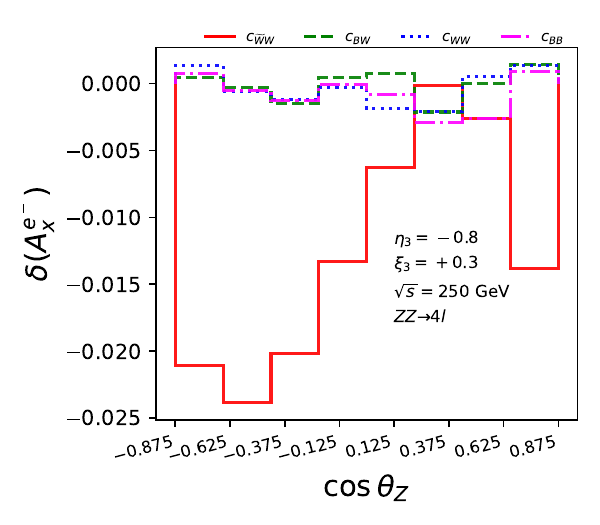}
     \includegraphics[width=0.325\textwidth]{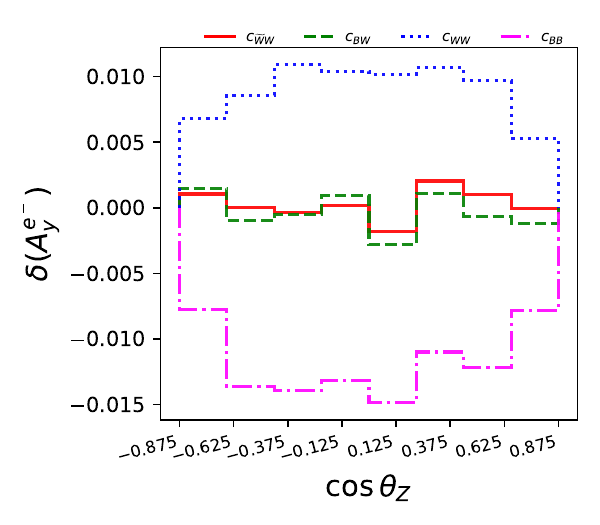}
     \includegraphics[width=0.325\textwidth]{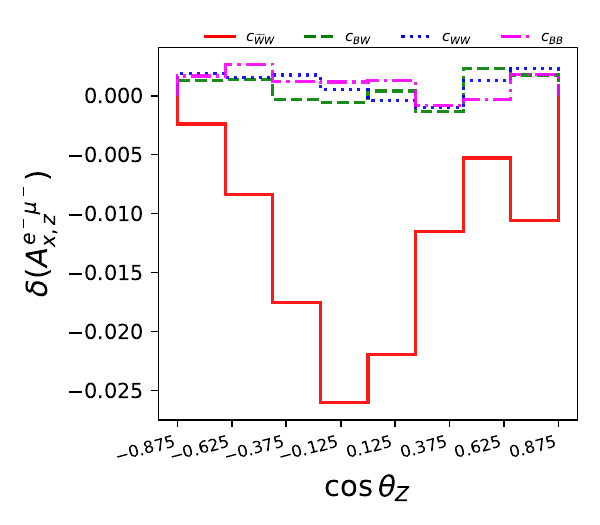}
     \caption{Relative change in asymmetries related to the CP-even~($P_x$) and CP-odd~($P_y$) polarization and correlation~($A_{x,z}$) where both the functions are CP-even. The distributions are shown for two set of beam polarization, $\eta_3,\xi_3=+0.8,-0.3$~(in the top row) and the flipped polarization~(in the bottom row) for four different anomalous couplings.  The analysis is done at $\sqrt{s}=250$~GeV, and at the rest frame of the mother particle.}
     \label{fig:asymm}
 \end{figure*}
The spin of a particle dictates the couplings with other fermions and bosons. These couplings are affected by the presence of higher-order operators; thus, understanding these couplings will serve as a window to probe deviations from the prediction of SM. The change in couplings may induce deviation in the values of various observables like cross~section, momenta, and angular distribution of final state particles. Some of the asymmetries constructed out of the angular distribution of final particles carry information on the spin of the mother particle in the form of polarization parameters. The polarization parameters are related to the dynamics of particle production; thus, these coefficients or spin-related observables can serve as a valuable tool for constraining the parameters of new physics. Generally, a particle with spin-$s$ provides $4s(s+1)$ polarization parameters, which are the independent parameters of the particle spin density matrix. For a spin-1 particle like $Z$ boson, the spin density matrix can be parameterized in Cartesian form as,
\begin{equation}
\label{eqn:sdm}
    P(\lambda,\lambda^\prime) = \frac{1}{3}\left[I + \frac{3}{2}\Vec{P}\cdot\Vec{S} + \sqrt{\frac{3}{2}}T_{ij}\{S_i,S_j\}\right],
\end{equation}
where $\Vec{P}$ and $T_{ij}$ are the vector and tensor polarization parameters of $Z$ boson respectively and $\Vec{S}$ is the spin-1 operator. To estimate them, one can calculate the parameters from the production density matrix, which is proportional to the product of the production amplitudes for different helicities of the $Z$ boson, $\rho(\lambda,\lambda^\prime) \propto M(\lambda)M^\star(\lambda^\prime)$, where $\lambda \in \{-,0,+\}$. Alternatively, one can constraint them at the decay level using the angular distribution of the decay products as in collider experiments. In this article, we focus on constructing polarization of $Z$ boson using the later methods. The angular distribution of fermions decayed from a spin-1 boson is written as~\cite{Boudjema:2009fz},
\begin{equation}
    \label{eqn:domegaone}
    \begin{split}
	   &\frac{1}{\sigma}\frac{d\sigma}{d\Omega} = \frac{3}{8\pi}\left[\left(\frac{2}{3}-(1-3\delta)\frac{T_{zz}}{\sqrt{6}}\right)+\alpha P_z\cos\theta \right.\\&+\left. \sqrt{\frac{3}{2}}(1-3\delta)T_{zz}\cos^2\theta +\left(\alpha P_x +2\sqrt{\frac{2}{3}}(1-3\delta)T_{xz}\cos\theta\right)\right.\\&\times\left.\sin\theta\cos\theta + \left(\alpha p_x +2\sqrt{\frac{2}{3}}(1-3\delta)T_{yz}\cos\theta\right)\sin\theta\sin\phi \right.\\&+\left.(1-3\delta)\sin^2\theta\left\{\left(\frac{T_{xx}-T_{yy}}{\sqrt{6}}\right)\cos(2\phi)\right.\right.\\&+\left.\left.\sqrt{\frac{2}{3}}T_{xy}\sin(2\phi)\right\}\right].
    \end{split}
\end{equation}
Here, $\theta,\phi$ are the polar and azimuthal angle of the final state fermion in the rest frame of the $Z$ boson with its would-be momentum along the $z$-direction and $x-z$ plane being the lab frame production plane. The parameters $\alpha$ and $\delta$ depend on the chiral couplings and ratio of the mass of final state fermions to the mass of resonance, such that at the limit of massless fermions, $\alpha\to \frac{R^2_f-L^2_f}{R^2_f+L^2_f}$ and $\delta \to 0$~\cite{Boudjema:2009fz}. At the decay level, one can calculate various asymmetries related to polarization and spin correlation parameters by taking the partial integration of Eq.~(\ref{eqn:domegaone}) w.r.t $\theta$ and $\phi$. For example, we can get $P_x$ from the left-right asymmetry as~\cite{Boudjema:2009fz},
\begin{equation}
    \begin{aligned}
        A_x &= \frac{1}{\sigma}\left[\int_{\theta=0}^\pi\int_{\phi=-\frac{\pi}{2}}^{\frac{\pi}{2}}\frac{d\sigma}{d\Omega}d\Omega - \int_{\theta=0}^\pi\int_{\phi=\frac{\pi}{2}}^{3\frac{\pi}{2}}\frac{d\sigma}       {d\Omega}d\Omega\right] \\
        &= \frac{3\alpha P_x}{4} = \frac{\sigma(f_x > 0) - \sigma(f_x < 0)}{\sigma(f_x > 0) + \sigma(f_x < 0)},
    \end{aligned}
\end{equation}
where $f_x = \sin\theta\cos\phi$. Likewise, the other angular functions used to construct the asymmetries are $f_y=\sin\theta\sin\phi$, $f_z=\cos\theta$, $f_{xy}=f_xf_y$, $f_{xz}=f_xf_z$, $f_{yz} = f_yf_z$, $f_{x^2-y^2} = f_x^2 - f_y^2$, and $f_{zz} = \sqrt{1-f_z^2}\left[3-4(1-f_z^2)\right]$. The asymmetries related to other polarization parameters are obtained similarly using the above function.\\\\ 
When two particles are co-produced, their spin is correlated owing to the conservation of angular momentum. In general, for two particles $A$ and $B$ with spin $s_A$ and $s_B$ respectively, there are $16s_As_B(s_A+1)(s_B+1)$ spin correlation parameters. The total rate for the production of two spin-1 particles followed by their decay, .i.e, $A\to aa^\prime$ and $ B\to bb^\prime$ can be written as~\cite{Rahaman:2021fcz},
 \begin{equation}
 \begin{aligned}
     \frac{1}{\sigma}\frac{d\sigma}{d\Omega_a d\Omega_b} = &\frac{9}{16\pi^2}\sum_{\lambda_A,\lambda_A^\prime,\lambda_B,\lambda_B^\prime}
     P_{AB}(\lambda_A,\lambda_A^\prime,\lambda_B,\lambda_B^\prime)\\&\Gamma_A(\lambda_A,\lambda_A^\prime)\Gamma_B(\lambda_B,\lambda_B^\prime).
 \end{aligned}
 \end{equation}
 $P_{AB}$ represents the joint polarization density matrix. Moreover, similarly, by using the respective form for the density matrix, one obtains the joint angular distribution of the decay products of $A$ and $B$. In the case of two spin-1 particles, there will be 64 spin correlation parameters, which can be obtained using different angular functions as~\cite{Rahaman:2021fcz}, 
 \begin{equation}
     \mathcal{A}_{ij}^{AB} = \frac{\sigma(f^a_if^b_j > 0) - \sigma(f^a_if^b_j < 0)}{\sigma(f^a_if^b_j > 0) + \sigma(f^a_if^b_j < 0)}.
 \end{equation} \\
 It is worth noting that the asymmetries defined above can be divided into CP-even and CP-odd observables. Three of the polarization asymmetries are CP-odd, and five are CP-even. In the case of spin correlations, there are 34 CP-even and 30 CP-odd asymmetries. In Fig.~\ref{fig:asymm}, we show the relative change of asymmetries value in presence of anomalous couplings by keeping one anomalous couplings to non-zero at a time while other are set to zero. The benchmark anomalous point is kept to 100~TeV$^{-2}$ for each new physics parameter. In particular, we show the variation for asymmetries~(polarization) related to the angular distribution of $f_x^{e^-}~(A_x^{e^-})$, $f_y^{e^-}~(A_y^{e^-})$ and correlation of function~$f_x^{e^-}f_z^{\mu^-}~(A_{x,z}^{e^-\mu^-})$. We perform this analysis at center-of-mass energy of $250$~GeV, with two set of initial beam polarization $(\eta_3,\xi_3) = (+0.8,-0.3)$~(shown in the top row) and the flipped polarization~(in the bottom row). In Fig.~\ref{fig:asymm}, we see that the variation in the asymmetries for CP-odd anomalous couplings, $c_i\in\{c_{WW},c_{BB},c_{BW}\}$ are minimal while the variation are significant for CP-even asymmetries like $A_x$, and $A_{x,z}^{e^-\mu^-}$, while it provides significant change in CP-odd asymmetries~($A_y^{e^-}$). Further, in case of CP-even asymmetries the sign of the relative change in asymmetries gets flipped in the case when the beam polarization is flipped. Thus, various asymmetries might act as an analyzer for the CP state of the new physics. The change in the asymmetries due to anomalous couplings are non-symmetric to the $\cos\theta_Z$ related to the $Z$ boson. E.g., in the case of asymmetries related to $A_{x,z}$, the maximal change from the SM value is seen for $-0.5<\cos\theta_Z < 0.5$ region, and in the case of $A_x$, the change follows a asymmetric distribution in the range $-1.0<\cos\theta_Z<+1.0$. This asymmetric change w.r.t $\cos\theta_Z$ allows one to bin the asymmetries in a specific range of $\cos\theta_Z$.\\\\ 
In the next section, we discuss the methodology and the limits obtained on various anomalous couplings. We will also emphasize the role of statistical and systematic errors in setting the bounds on anomalous couplings.

\section{Probe of anomalous couplings}
\label{sec:anomalous}
This section presents a comprehensive methodology for obtaining bounds on anomalous couplings, denoted by $c_i$, in the context of $ZZ$ and $Z\gamma$ processes. Our approach involves dividing all the observables into eight bins of $\cos\theta_Z$, where $\theta_Z$ represents the production angle of the $Z$ boson in the laboratory frame. This allows us to increase the number of observables and better constrain the anomalous couplings. This strategy has the advantage of fine-tuning the distinction between the signal resulting from new physics and the SM forecast. The ability to detect and eliminate differences between the SM predictions and experimental data is made possible by the growth in observables, which in turn allows us to impose stricter limits on anomalous couplings. The event simulations are carried out using the Monte Carlo method implemented in {\tt MadGraph5$\_$aMC$@$NLO}~\cite{Alwall:2014hca}, which allows us to simulate the production and decay of $ZZ$ and $Z\gamma$ bosons in the presence of anomalous couplings. The resulting simulated events are then analyzed to extract the relevant observables, which are divided into the eight bins mentioned above of $\cos\theta_Z$. \\\\
Next, we fit the values of the observables in each bin to derive a semi-analytical expression for all observables as a function of the anomalous couplings. This enables us to establish the relationship between the observables and the anomalous couplings and to derive the most likely values of the couplings that best fit the data. In presence of anomalous couplings, $c_i$, the cross~section is fitted using the following parameterization,
\begin{equation}
\label{eqn:fitxsec}
    \begin{aligned}
        \sigma &= \sigma_0 + \sigma_ic_i +\sum_{j\neq k}\sigma_{jk}c_jc_k+ \sum_{l=1}^4\sigma_{ll}c_l^2,\\
        & i = c_{\widetilde{B}W}, j,k \in \{c_{BW},c_{WW},c_{BB}\}, l = i \cup j.
    \end{aligned}
\end{equation}
The numerator and denominator are fitted independently for asymmetries and then used as
\begin{equation}
    \mathcal{A}_i(\{c_i\}) = \frac{\Delta\sigma_{\mathcal{A}_i}(\{c_i\})}{\sigma(\{c_i\})}.
 \end{equation}
The denominator is the cross~section which is fitted as in Eq.~(\ref{eqn:fitxsec}), and the numerator of CP-odd asymmetries is fitted as,
\begin{equation}
\label{eqn:cpodd}
    \begin{aligned}
         \Delta\sigma_{\mathcal{A}_i}(\{c_i\}) &= \sum_i \sigma_i c_i + \sum_i\sigma_{1i}c_{\widetilde{B}W}c_i,\\
         & i \in \{c_{BW},c_{WW},c_{BB}\}.
    \end{aligned}
\end{equation}
\begin{figure*}[!htb]
    \centering
    \includegraphics[width=0.49\textwidth]{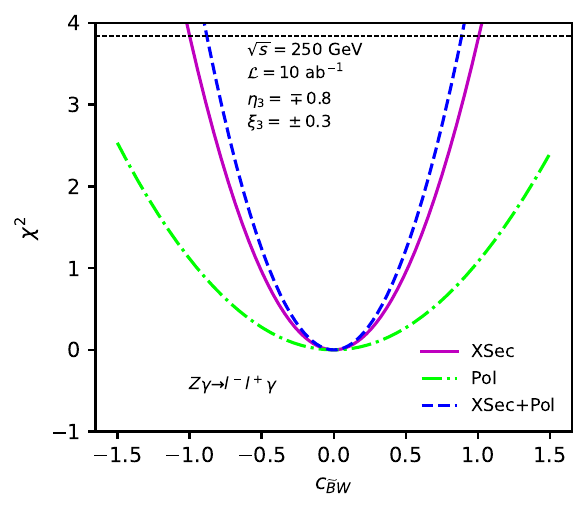}
    \includegraphics[width=0.49\textwidth]{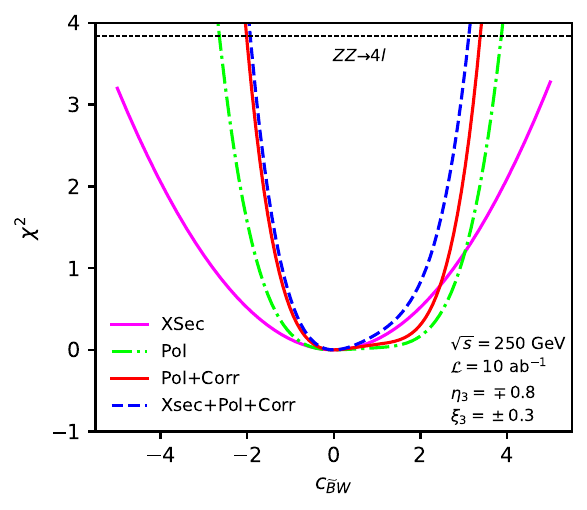}
    \includegraphics[width=0.49\textwidth]{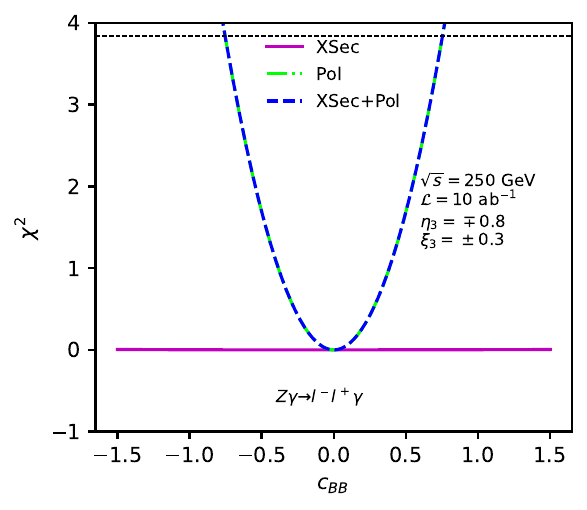}
    \includegraphics[width=0.49\textwidth]{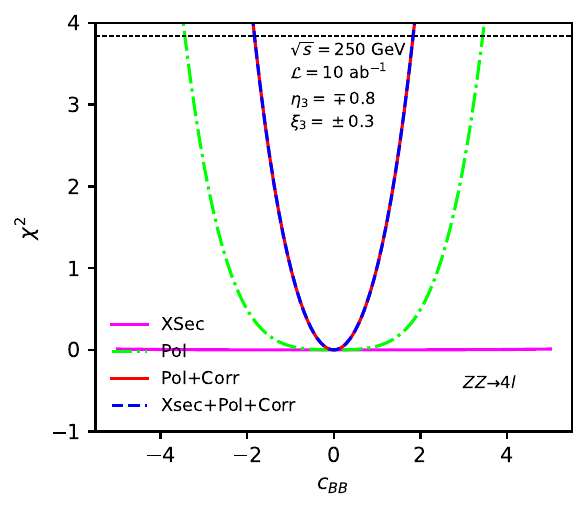}
    \caption{One dimensional $\chi^2$ for cross-section, polarizations, combinations of polarization and spin-spin correlations and combinations of all observables as a function of one anomalous couplings at a time for a $ZZ\to 4l$ and $Z\gamma\to l^-l^+\gamma$ process. The plots are obtained at $\sqrt{s}=250$ GeV, $\mathcal{L}=10$ ab$^{-1}$ and zero systematic errors. The horizontal line at $\chi^2=3.84$ corresponds to the limits of anomalous couplings at 95$\%$ confidence level~(CL).}
    \label{fig:onechi}
\end{figure*}
When dealing with events with two sets of beam polarization, merging the different sets at the $\chi^2$ level yields more stringent constraints~\cite{Rahaman:2019mnz}. The combined $\chi^2$ at different set of beam polarization~($\pm\eta_3,\mp\xi_3$), and using different observable $\mathcal{O}$ for a given value of anomalous couplings $c_i$ is defined as,
\begin{equation}
\begin{aligned}
    &\chi^2(\mathcal{O},c_i,\pm\eta_3,\mp\xi_3) = \\ &\sum_{l,k} \left[\left(\frac{\mathscr{O}_k^l(c_i,+\eta_3,-\xi_3) - \mathscr{O}_k^l(0,+\eta_3,-\xi_3)}{\delta \mathscr{O}^l_k(0,+\eta_3,-\xi_3)}\right)^2 \right.\\ &\left.+  \left(\frac{\mathscr{O}_k^l(c_i,-\eta_3,+\xi_3) - \mathscr{O}_k^l(0,-\eta_3,+\xi_3)}{\delta \mathscr{O}^l_k(0,-\eta_3,+\xi_3)}\right)^2 \right],
\end{aligned}
\label{eqn:chisum}
\end{equation}
where $k,l$ runs over all bins and observables separately, and $\delta\mathcal{O} = \sqrt{(\delta\mathcal{O}_{\text{stat}})^2 + (\delta\mathcal{O}_{\text{syst}})^2}$ is the estimated error in $\mathcal{O}$. For cross~section $\sigma$, the estimated error is given by,
\begin{equation}
    \delta\sigma = \sqrt{\frac{\sigma}{\mathcal{L}}+(\epsilon_\sigma\sigma)^2},
    \label{eqn:xsecstat}
\end{equation}
and if the observable is asymmetry, the error is given by,
\begin{equation}
    \delta\mathcal{A} = \sqrt{\frac{1-\mathcal{A}^2}{\mathcal{L}\sigma}+\epsilon_\mathcal{A}^2}.
\end{equation}
Here, $\epsilon_\mathcal{A},\epsilon_\sigma$ are the systematic error in asymmetry and cross~section, respectively, and $\mathcal{L}$ is the integrated luminosity. The chi-squared analysis provides complete information on the sensitivity of observables along with the constraint on anomalous couplings.
\begin{figure*}[!htb]
    \centering
    \includegraphics[width=0.49\textwidth]{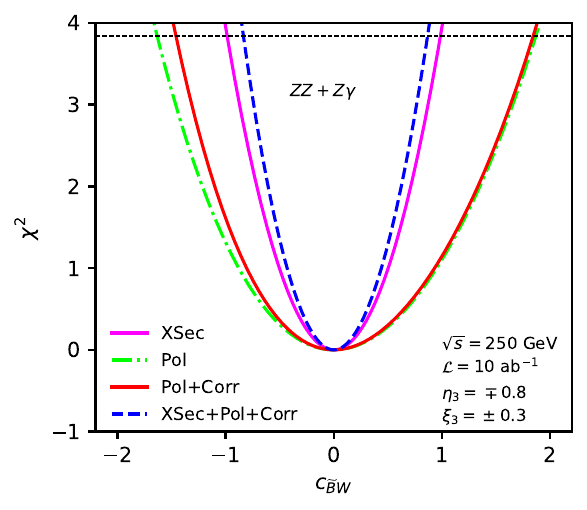}
    \includegraphics[width=0.49\textwidth]{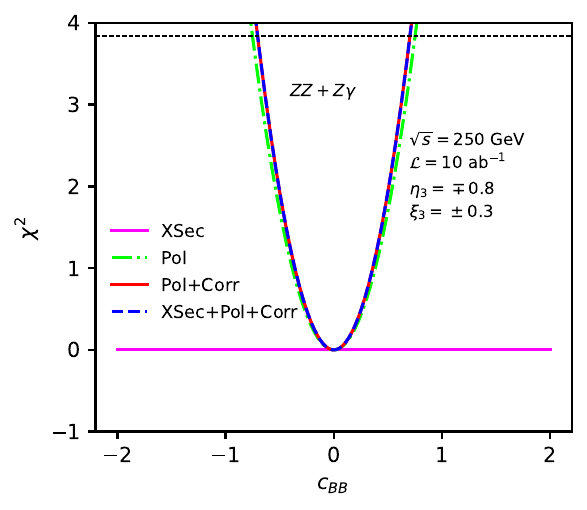}
    \caption{\label{fig:Onedim}One dimensional combined $\chi^2$ for different sets of observables as a function of anomalous couplings~($c_i\in\{c_{\widetilde{B}W},c_{BW}\}$). The horizontal line at $\chi^2=3.84$ denotes limits at $95\%$ CL. The analysis is done with zero systematic error.}
\end{figure*}
The analysis are performed at $\sqrt{s}=250$~GeV with different values of integrated luminosity,
\begin{equation}
\label{eqn:lumi}
\mathcal{L}\in\{0.1~\text{ab}^{-1}, 0.3~\text{ab}^{-1}, 1~\text{ab}^{-1},3~\text{ab}^{^-1},10~\text{ab}^{-1}\}.
\end{equation}
The SM cross~sections for $ZZ\to e^-e^+\mu^-\mu^+$ and $Z\gamma\to l^-l^+\gamma$ process at $\sqrt{s}=250$ GeV with two set of initial beam polarization of ($\pm0.8,\mp0.3$) are,
\begin{equation}
    \begin{aligned}
        \sigma_{ZZ}(+0.8,-0.3) = 2.15~\text{fb},& \\
        \sigma_{ZZ}(-0.8,+0.3) = 3.41~\text{fb},&\\
        \sigma_{Z\gamma}(+0.8,-0.3) = 607.62~\text{fb},&\\
        \sigma_{Z\gamma}(-0.8,+0.3) = 765.83~\text{fb}.&
    \end{aligned}
    \label{eqn:xsec}
\end{equation}
Given the cross~section as in Eq.~(\ref{eqn:xsec}) and luminosity in Eq.~(\ref{eqn:lumi}), the relative statistical error as obtained from Eq.~(\ref{eqn:xsecstat}) are,
\begin{equation}
    \begin{aligned}
    \frac{\delta\sigma}{\sigma}(+0.8,-0.3)\vert_{ZZ} &= \{6.81\%,3.93\%,2.15\%,1.24\%,0.68\%\},\\
    \frac{\delta\sigma}{\sigma}(-0.8,+0.3)\vert_{ZZ} &= \{5.40\%,3.12\%,1.71\%,0.98\%,0.54\%\},\\
    \frac{\delta\sigma}{\sigma}(+0.8,-0.3)\vert_{Z\gamma} &= \{0.40\%,0.23\%,0.12\%,0.07\%,0.04\%\},\\
    \frac{\delta\sigma}{\sigma}(-0.8,+0.3)\vert_{Z\gamma}&=\{0.36\%,0.20\%,0.11\%,0.06\%,0.03\%\}.
    \end{aligned}
    \label{eqn:error}
\end{equation}
\begin{figure*}[!htb]
    \centering
    \includegraphics[width=0.49\textwidth]{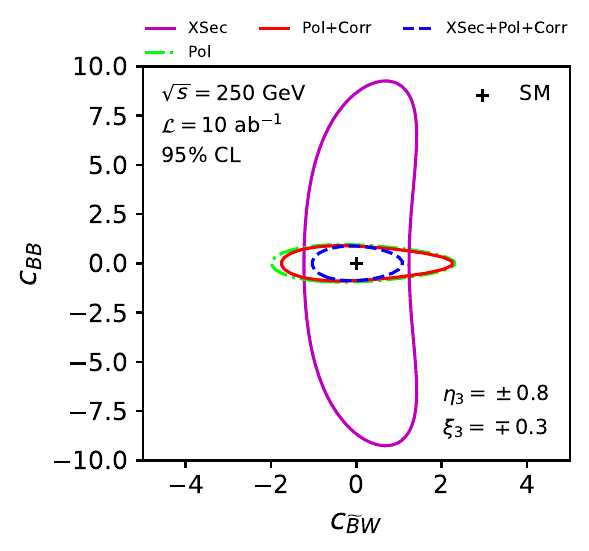}
    \includegraphics[width=0.49\textwidth]{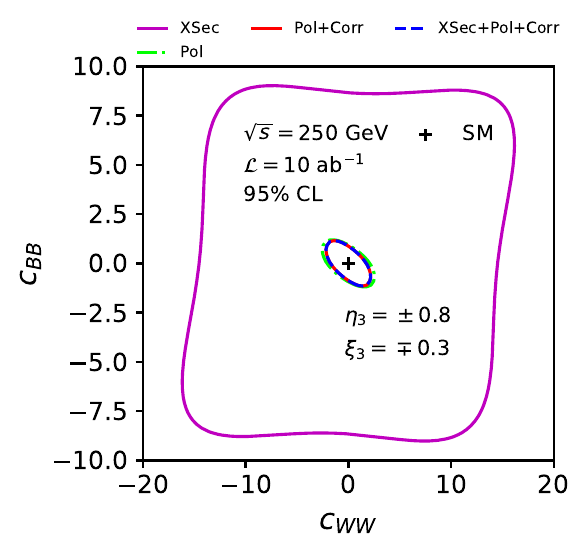}
    \caption{2-D 95$\%$ CL contours of $\chi^2$ for cross-section, polarizations, spin correlations and their combinations as a function of two anomalous couplings at a time for combined $ZZ$ and $Z\gamma$process. The analysis is done at center of mass energy $\sqrt{s}=250$ GeV, integrated luminosity $\mathcal{L}=10$ ab$^{-1}$ and systematic error are chosen to be zero. The cross in the center represents SM points.}
    \label{fig:twodim}
\end{figure*}
The above statistical error is given for a unbinned events, while for the binned~(eight bin) case the statistical error increases depending on the rate. On top of the statistical error given in the above equation, we add systematic error given as,
\begin{equation}
\label{eqn:syst}
    (\epsilon_\sigma,\epsilon_\mathcal{A}) = \{(0,0),(0.5\%,0.25\%),(2\%,1\%)\}.
\end{equation}
In the case of $ZZ\to4l$ events, the estimated error is primarily dominated by the statistical component within the regime of systematic error given by Eq.~(\ref{eqn:syst}) due to the low branching ratio of the full lepton channel. It implies that the sensitivity of various observables would increase with luminosity, while for the case of $Z\gamma$ process the net error saturates for given non-zero value of systematic error.   \\\\
The figures presented in Fig.~\ref{fig:onechi} depict the $\chi^2$ analysis performed for various sets of observables as a function of one anomalous coupling at a time for two different final topologies: $4l$ and $l^-l^+\gamma$. The analysis is carried with an integrated luminosity of $\mathcal{L}=10$ ab$^{-1}$, and with systematic errors chosen to be zero. The beams are polarized with a degree of polarization $\eta_3 = \pm 0.8$ and $\xi_3 = \mp 0.3$.\\
Each panel of Fig.~\ref{fig:onechi} illustrates that the dominant contribution to constraining anomalous couplings arises from the spin-related observables, which are combinations of the polarization and spin-spin correlation parameters. For CP-odd couplings in the bottom row of the Fig.~\ref{fig:onechi}, the limits set by these spin-related observables remain unaffected by the presence of a cross~section. This is because, in the case of CP-even observables like cross~section, the contribution from CP-odd couplings only arises at the $1/\Lambda^8$ term, whose contribution is sub-dominant compared to the linear contribution~($1/\Lambda^4$) from CP-even. In the presence of one anomalous coupling ($c_i$), the CP-even observables can be parameterized by Eq.~(\ref{eqn:xsec1}), 
\begin{equation}
\label{eqn:xsec1}
    \sigma(c_i) = \sigma_0 + \sigma_i c_i + \sigma_{ii}c_{i}^2,
\end{equation}
where $\sigma_0$ corresponds to the SM value. Therefore, the linear term in the cross~section is absent for CP-odd $c_i$, leading to a negligible contribution. It is also evident from Fig.~\ref{fig:onechi} that the $Z\gamma$ process imposes a tighter constraint than the $ZZ$ process in the case of CP-odd coupling, $C_{BB}$ through combination of different polarization asymmetries. However, for the CP-even case, the cross~section provides tighter constraint than the polarization asymmetries in case of $Z\gamma$ process, while the limits for the same anomalous coupling from $ZZ$ process is bit relaxed.
\begin{figure*}[!htb]
    \centering
    \includegraphics[width=0.32\textwidth]{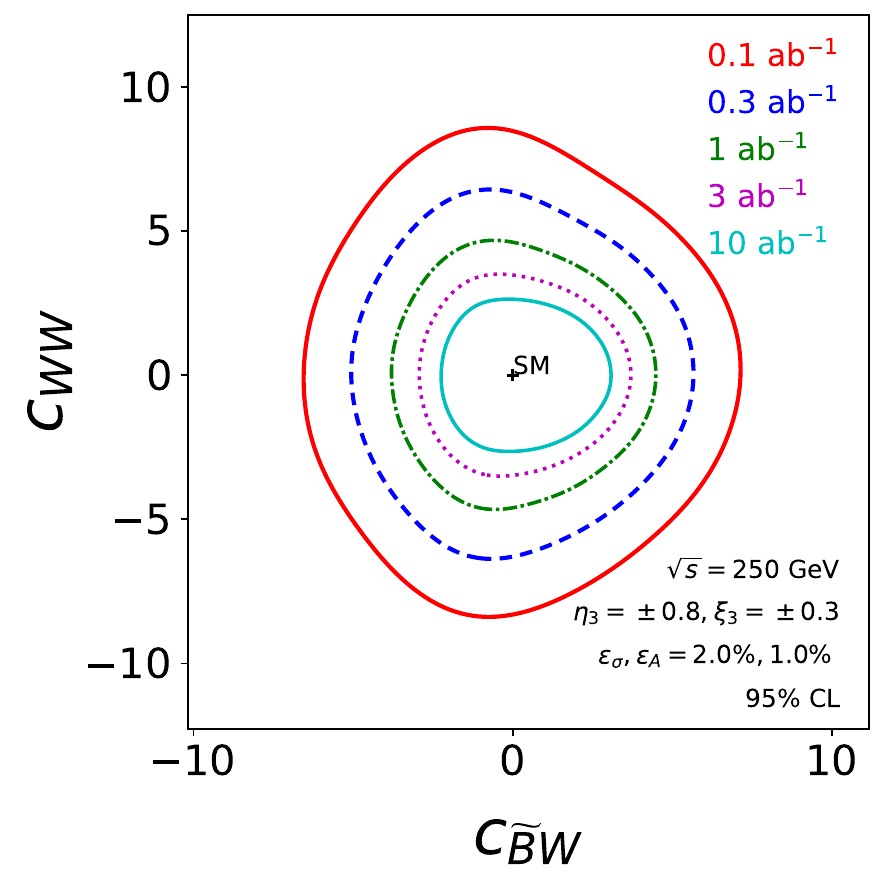}
    \includegraphics[width=0.32\textwidth]{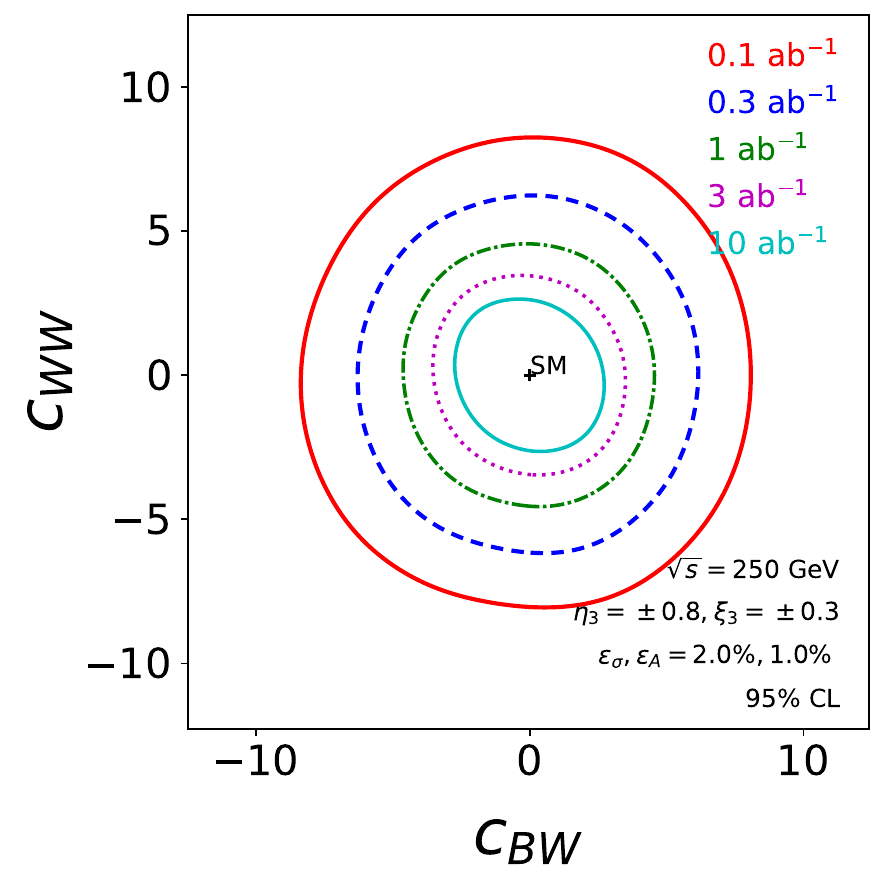}
    \includegraphics[width=0.32\textwidth]{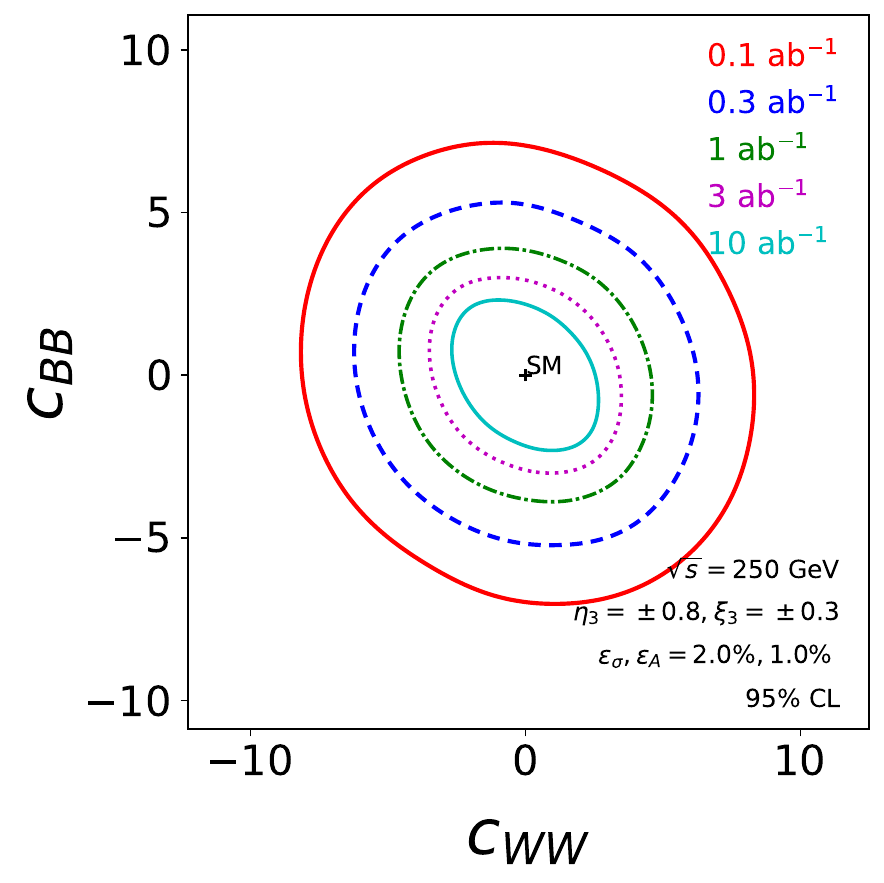}
    \caption{Two dimensional marginalized projections of the anomalous couplings $c_i$~(TeV$^{-4}$) at 95$\%$ CL obtained using MCMC global fits for a conservative systematic error $(2.0\%,1.0\%)$ and different values of integrated luminosities.}
    \label{fig:twodlumi}
\end{figure*}
\begin{figure*}[!htb]
    \centering    
    \includegraphics[width=0.32\textwidth]{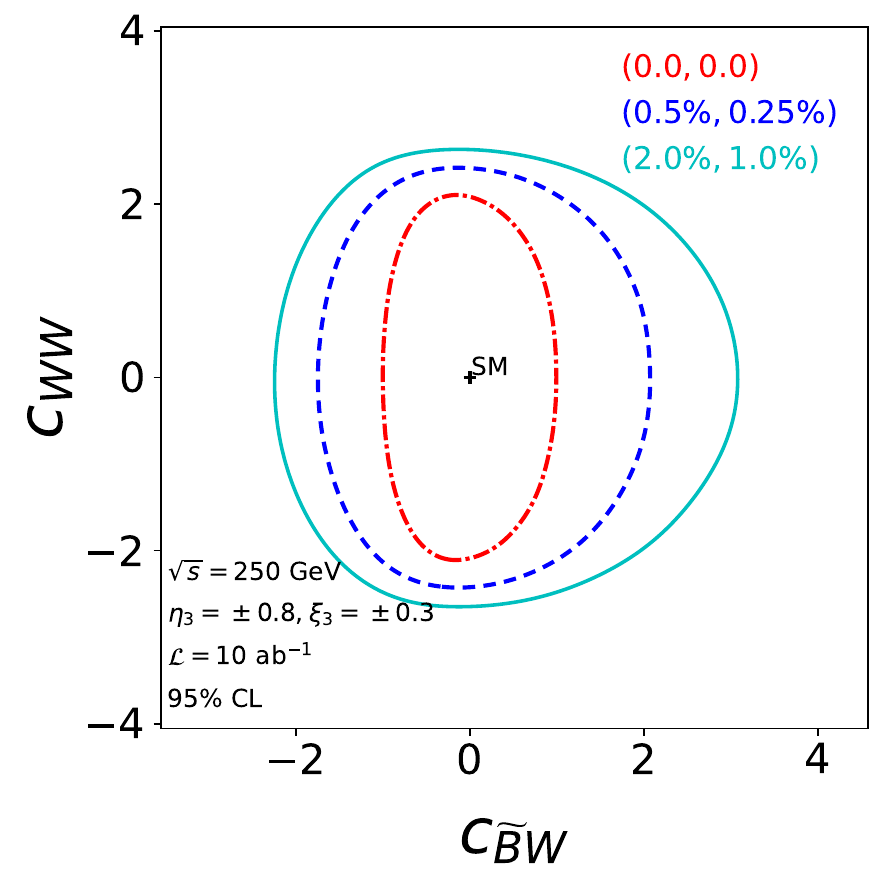}
    \includegraphics[width=0.32\textwidth]{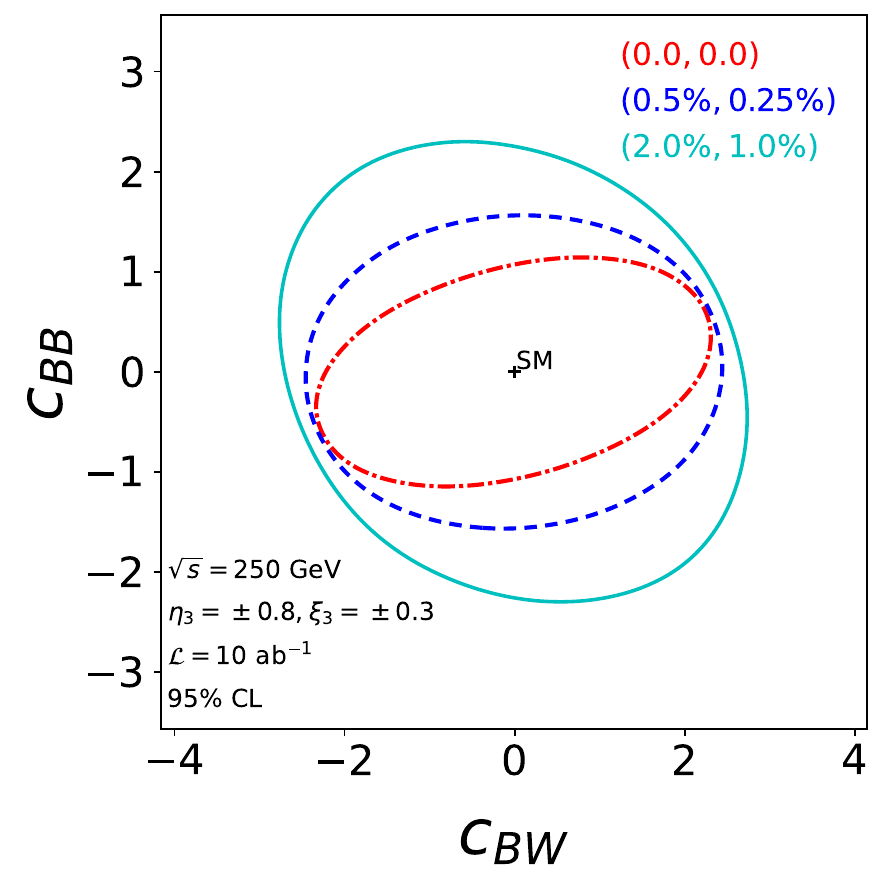}
    \includegraphics[width=0.32\textwidth]{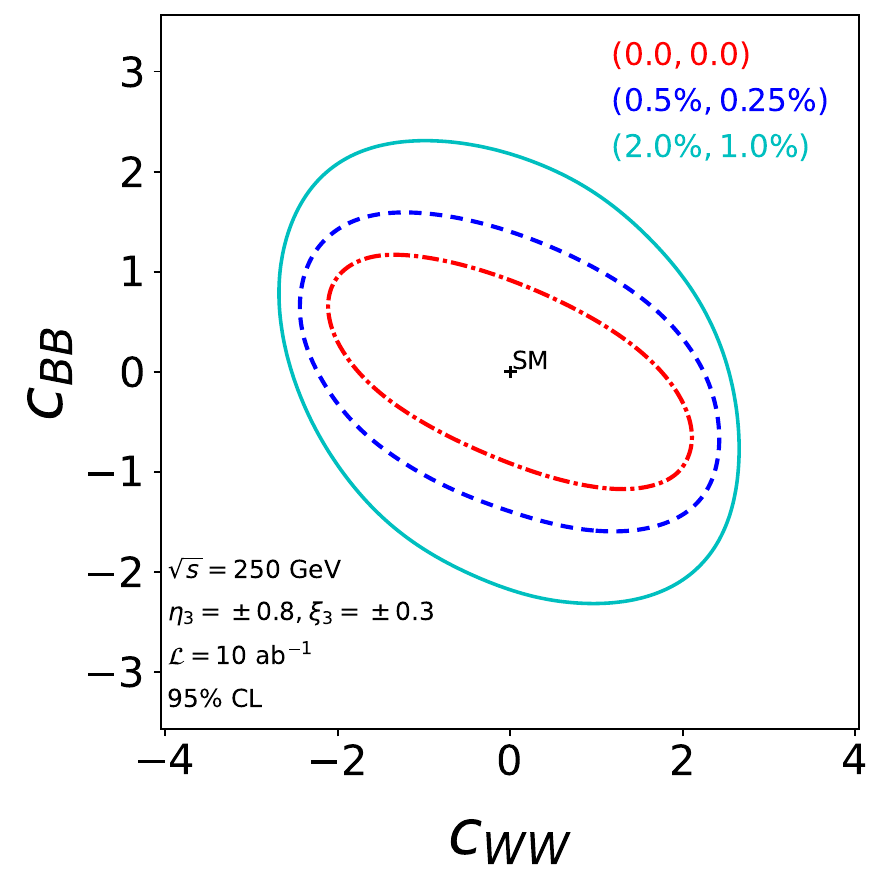}
    \caption{Two dimensional marginalized projections of the maximally correlated parameters $c_i$~(TeV$^{-4}$) at 95$\%$ CL obtained using MCMC global fits for a set of systematic error and integrated luminosity $\mathcal{L}=10$ ab$^{-1}$.}
    \label{fig:twodsyst}
\end{figure*}
Next, we combine the two processes at the level of $\chi^2$ as,
\begin{equation}
    \label{eqn:comchipro}
    \begin{aligned}
        \chi^2_{\text{Tot}}(\{c_i\},\pm\eta_3,\mp \xi_3) &= \chi^2_{ZZ}(\{c_i\},\pm\eta_3,\mp \xi_3) \\&+ \chi^2_{Z\gamma}(\{c_i\},\pm\eta_3,\mp\xi_3),
    \end{aligned}
\end{equation}
where each $\chi^2$ is obtained following Eq.~(\ref{eqn:chisum}) and on obtaining the above defined $\chi^2$, we have consider the all other leptonic decay channel for $ZZ$ process. The resultant one and two-dimensional $\chi^2$ distribution for a set of observables as a function of one and two anomalous couplings at a time are shown in Fig.~\ref{fig:Onedim} and Fig.~\ref{fig:twodim}, respectively. The hierarchy of observables on setting the bounds on anomalous couplings is the same as described above for individual topology. For CP-odd case, the cross~section provides the least contribution to the $\chi^2$, and combinations of polarization and spin correlation asymmetries predominantly set the bounds. For the CP-even case, the cross~section contributes the most to $\chi^2$ alone. The one parameter limits at 95$\%$ confidence level~(CL) obtained from the combined analysis are listed in Table~\ref{tab:Lim0ne}, where it is evident that the $Z\gamma$ channel dominates the limits on most of the anomalous couplings, except for $c_{BW}$.
\begin{table}[!h]
    \centering
    \caption{\label{tab:Lim0ne}One parameter limits on anomalous couplings~($c_i$) at 95$\%$ CL obtained for $ZZ\to 4l$, $Z\gamma\to 2l\gamma$ and their combinations. The limits are obtained at center of mass energy $\sqrt{s}=250$ GeV, integrated luminosity $\mathcal{L} = $ 10 ab$^{-1}$ and zero systematic.}
    \begin{tabular}{llll}
    \hline \hline
    \multirow{2}{*}{Parameters~($c_i$)}&\multicolumn{3}{c}{Limits~(TeV$^{-4}$)}\\
    \cline{2-4}
     &$Z\gamma\to 2l\gamma $&$ZZ\to 4l$&$ZZ+Z\gamma$\\   \hline
     $c_{\widetilde{B}W}/\Lambda^4$&$[-0.88,+0.88]$&$[-1.93,+3.11]$&$[-0.83,+0.86]$\\
     $c_{BW}/\Lambda^4$&$[-2.89,+2.89]$&$[-2.40,+2.40]$&$[-1.97,+1.97]$\\
     $c_{WW}\Lambda^4$&$[-1.75,+1.75]$&$[-2.17,+2.17]$&$[-1,42,+1.42]$\\
     $c_{BB}/\Lambda^4$&$[-0.75,+0.75]$&$[-1.83,+1.83]$&$[-0.70,+0.70]$\\
     \hline\hline
    \end{tabular}
\end{table}
The one parameter limits in the current work~(ILC $@$ 10~ab$^{-1}$) are diluted for $Z\gamma$ process, while the limits are tighter for $ZZ$ process, in contrast to the experimental limit~(LHC $@$ 36.1~fb$^{-1}$) listed in Table~\ref{tab:const}. Our current study examines the ILC potential to probe dim-8 bosonic operators without directly comparing it to existing colliders. Notably, the LHC boasts a higher collision energy~($\approx$ 2 TeV) compared to the ILC~(250 GeV). This makes the anomalous vertex more sensitive to the higher energy reach of LHC since the vertex created by dim-8 operators comprises the terms reliant on the momentum transfer normalized by the electroweak scale. Besides the energy domination at LHC, the cross-section of $Z\gamma \to \nu\bar{\nu}\gamma$ is $\approx 25$ times larger at ILC than it is for $Z\gamma \to l^-l^+\gamma$, , while for $ZZ\to4l$ process the cross~section is $\approx 9$ times higher in LHC than at ILC. Nevertheless, the significant number of observables used in our analysis try to compensate for the low energy and cross~section, which is visible in Fig.~\ref{fig:Onedim} where the limits on anomalous couplings become tighter on using polarizations and spin correlations. A recent study~\cite{Rahaman:2016pqj} achieved a comparable limit on anomalous couplings at greater energy $\sqrt{s}=500$~GeV, using fewer observables (polarizations of $Z$ bosons). Similar study~\cite{Rahaman:2018ujg} at LHC at 13~TeV using polarizations of $Z$ boson in final $4l$ events obtained a tighter constraint on anomalous couplings. The higher collision energy compensates for the higher number of observables.\\
In case of two dimensional contours shown in Fig.~\ref{fig:twodim}, the cross~section in presence of two anomalous couplings~($c_i,c_j$) can be parameterized as,
\begin{equation}
    \label{eqn:xsectwo}
    \begin{aligned}
        \sigma_i(c_i,c_j) = &\sigma_0 + \sigma_i c_i + \sigma_{ij}c_ic_j + c_{ii}c_i^2+ c_{jj}c_j^2, \\ &c_i = c_{\widetilde{B}W}, c_j \in \{c_{BW},c_{WW},c_{BB}\}.
    \end{aligned}
\end{equation}
As mentioned earlier, when both parameters are CP-odd, the linear term vanishes; in the case of one CP-odd and CP-even parameter, the interference between anomalous parts vanishes. The shape described by Eq.(\ref{eqn:xsectwo}) can be seen in the left panel of Fig.~\ref{fig:twodim}, which contains one CP-even and one CP-odd coupling, resulting in tighter constraints on the x-axis than the y-axis. As in the one-parameter case, the spin-related observables have the maximal contribution in simultaneously constraining the two anomalous couplings. The results are provided for the rest of the analysis by combining the two processes as defined in Eq.~(\ref{eqn:comchipro}).
\begin{table*}
    \caption{\label{tab:95bcieft}List of posterior 95$\%$ Bayesian Confidence Interval~(BCI) of anomalous couplings for $\sqrt{s} = 250$ GeV with beam polarization~($\eta_3,\xi_3$) = ($\pm0.8,\pm0.3$) from MCMC global fits at different value of luminosity and systematic error. }
    \centering
    \renewcommand{\arraystretch}{1.4}
    \begin{tabular}{llllll}\\\hline\hline    
         $\mathcal{L}$~(fb$^{-1}$)&\quad\quad$(\epsilon_\sigma,\epsilon_A)$&\quad\quad$c_{\widetilde{B}W}$~ (TeV$^{-4}$)&\quad\quad$c_{BW}$~(TeV$^{-4}$)&\quad\quad$c_{WW}$~(TeV$^{-4}$)&\quad\quad$c_{BB}$~(TeV$^{-4}$)  \\ \hline         
         &\quad(0.0,0.0)&\quad\quad$[-4.81,+5.11]$&\quad\quad$[-6.83,+6.59]$&\quad\quad$[-6.64,+6.73]$&\quad\quad$[-5.31,+5.38]$\\
         100&\quad($0.5\%$,$0.25\%$)&\quad\quad$[-4.88,+5.21]$&\quad\quad$[-6.83,+6.60]$&\quad\quad$[-6.65,+6.74]$&\quad\quad$[-5.36,+5.42]$\\
         &\quad($2.0\%$,$1.0\%$)&\quad\quad$[-5.28,+5.76]$&\quad\quad$[-6.87,+6.63]$&\quad\quad$[-6.67,+6.79]$&\quad\quad$[-5.71,+5.81]$\\\hline
         
         &\quad(0.0,0.0)&\quad\quad$[-3.33,+3.51]$&\quad\quad$[-5.14,+5.02]$&\quad\quad$[-5.01,+5.02]$&\quad\quad$[-3.63,+3.64]$\\
         300&\quad($0.5\%$,$0.25\%$)&\quad\quad$[-3.51,+3.76]$&\quad\quad$[-5.14,+5.03]$&\quad\quad$[-5.04,+5.05]$&\quad\quad$[-3.73,+3.75]$\\
         &\quad($2.0\%$,$1.0\%$)&\quad\quad$[-4.08,+4.61]$&\quad\quad$[-5.19,+5.06]$&\quad\quad$[-5.11,+5.13]$&\quad\quad$[-4.28,+4.33]$\\\hline
         
          &\quad(0.0,0.0)&\quad\quad$[-2.18,+2.22]$&\quad\quad$[-3.75,+3.69]$&\quad\quad$[-3.60,+3.60]$&\quad\quad$[-2.31,+2.30]$\\
         1000&\quad($0.5\%$,$0.25\%$)&\quad\quad$[-2.49,+2.68]$&\quad\quad$[-3.76,+3.69]$&\quad\quad$[-3.67,+3.67]$&\quad\quad$[-2.49,+2.47]$\\
         &\quad($2.0\%$,$1.0\%$)&\quad\quad$[-3.08,+3.69]$&\quad\quad$[-3.83,+3.75]$&\quad\quad$[-3.77,+3.77]$&\quad\quad$[-3.19,+3.21]$\\\hline         
          &\quad(0.0,0.0)&\quad\quad$[-1.41,+1.38]$&\quad\quad$[-2.75,+2.72]$&\quad\quad$[-2.59,+2.58]$&\quad\quad$[-1.50,+1.49]$\\
         3000&\quad($0.5\%$,$0.25\%$)&\quad\quad$[-1.88,+2.11]$&\quad\quad$[-2.79,+2.75]$&\quad\quad$[-2.74,+2.73]$&\quad\quad$[-1.76,+1.75]$\\
         &\quad($2.0\%$,$1.0\%$)&\quad\quad$[-2.47,+2.91]$&\quad\quad$[-2.93,+2.88]$&\quad\quad$[-2.87,+2.86]$&\quad\quad$[-2.47,+2.47]$\\\hline
         
          &\quad(0.0,0.0)&\quad\quad$[-0.83,+0.84]$&\quad\quad$[-1.89,+1.87]$&\quad\quad$[-1.71,+1.70]$&\quad\quad$[-0.92,+0.92]$\\
         10000&\quad($0.5\%$,$0.25\%$)&\quad\quad$[-1.48,+1.66]$&\quad\quad$[-2.01,+1.99]$&\quad\quad$[-1.99,+1.98]$&\quad\quad$[-1.28,+1.28]$\\
         &\quad($2.0\%$,$1.0\%$)&\quad\quad$[-1.89,+2.48]$&\quad\quad$[-2.27,+2.48]$&\quad\quad$[-2.19,+2.18]$&\quad\quad$[-1.90,+1.90]$\\
                 \hline\hline
    \end{tabular}    
    \renewcommand{\arraystretch}{1.4}
\end{table*}
We conclude our analysis by performing a Markov Chain Monte Carlo (MCMC) computation using two sets of observables for two different polarized beams to derive simultaneous limits on all anomalous couplings ($c_i$). We aim to determine the likelihood of a given point $\bold{c}\in\{c_i,\pm\eta_3,\mp\xi_3\}$ in the parameter space. This is achieved by defining the likelihood function as the product of exponential functions of the $\chi^2$ function evaluated at $\bold{c}$ as,
\begin{equation}
\mathscr{L}(\bold{c}) = \prod_{i,j} \exp\left[-\frac{\chi^2(\bold{c})}{2}\right],
\end{equation}
where $\chi^2$ is defined in Eq.~(\ref{eqn:chisum}), and the indices $i$ and $j$ respectively run over the list of all bins and observables, including cross~sections, polarizations, and spin correlation asymmetries. We perform this analysis for a set of luminosities given in Eq.~(\ref{eqn:lumi}), as well as systematic error given in Eq.~(\ref{eqn:syst}). The MCMC implementation utilizes the Metropolis-Hastings algorithm. The marginalized limits on the Wilson's coefficient $c_i$ at $95\%$ CL is listed in Table~\ref{tab:95bcieft} for different values of integrated luminosity and systematic errors.\\\\
Fig.~\ref{fig:twodlumi} illustrates the two-dimensional marginalized projections at a 95$\%$ CL for the anomalous couplings~($c_i$) obtained from the MCMC global fits for a systematic error of 
$(2\%,1\%)$ and different luminosity values listed in Eq.~(\ref{eqn:lumi}). On increasing the luminosity from $0.1$~fb$^{-1}$ to $10$~fb$^{-1}$, the limits on each anomalous couplings gets better by an approximate factor of 3. For a case of zero systematic, the limits gets better by a factor of~$\approx$ 6 on increasing the luminosity by a factor of 100 while in case of a moderate systematic error~$(0.5\%,0.25\%)$, the limits gets tightened by nearly a factor of 3.4 with the same range of luminosity. Thus, the results points that the reduction of systematic errors becomes very important for the probe of the aNTGC. 
 \\\\
To comprehensively comprehend the impact of systematic error on constraining anomalous couplings, we analyze two-dimensional marginalized projections of anomalous couplings with a luminosity of $\mathcal{L}=10$~ab$^{-1}$~(see Fig.~\ref{fig:twodsyst}). In middle panel which depicts a pair plot of $(c_{BW},c_{BB})$, it is observed that the systematic errors have a minor impact on the marginalized limits of $c_{BW}$ coupling. It is due to a large statistical error associated with the $ZZ\to4l$ process, and this process predominantly sets the bounds on $c_{BW}$ through its large number of spin-related observables. The improvement on the limits of $c_{WW}$ is $\approx 1.28$, while for $c_{\widetilde{B}W}$, and $c_{BB}$ the limits improved by a factor of 2.3 and 2.1, respectively.\\
The marginalized limits suggest that the higher luminosity and lower systematic errors becomes of paramount importance to probe the non-standard triple gauge boson couplings in future lepton collider like ILC. The current study involved the leptonic decay of $Z$ boson, which has a low branching ratio. Analysis of the other remaining channels becomes necessary to increase the sensitivity of different polarization and spin correlation observables to anomalous couplings. In the case of hadronic decay, final jet flavor tagging becomes essential to reconstruct the parity odd asymmetries. In our previous studies~\cite{Subba:2022czw,Subba:2023rpm}, we have shown the effectiveness of machine learning techniques to tag the jets initiated by light quarks. A similar study of the semi-leptonic and hadronic decay of $Z$ bosons will be reported elsewhere.
\begin{figure*}[!htb]
    \centering
    \includegraphics[width=0.32\textwidth]{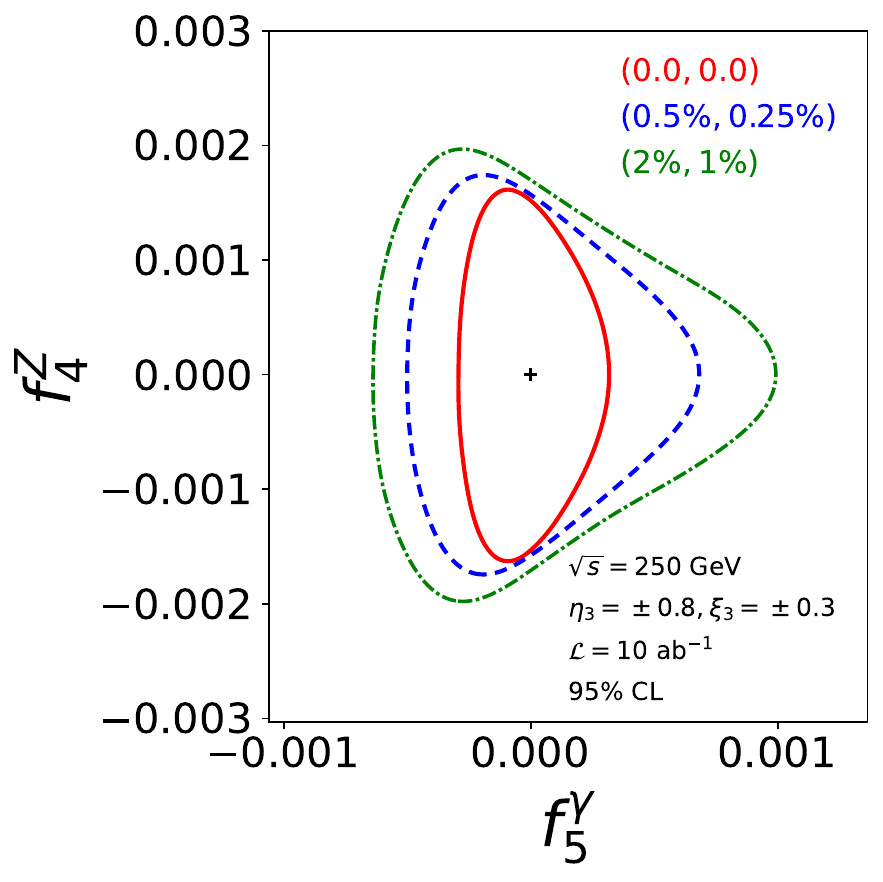}
    \includegraphics[width=0.32\textwidth]{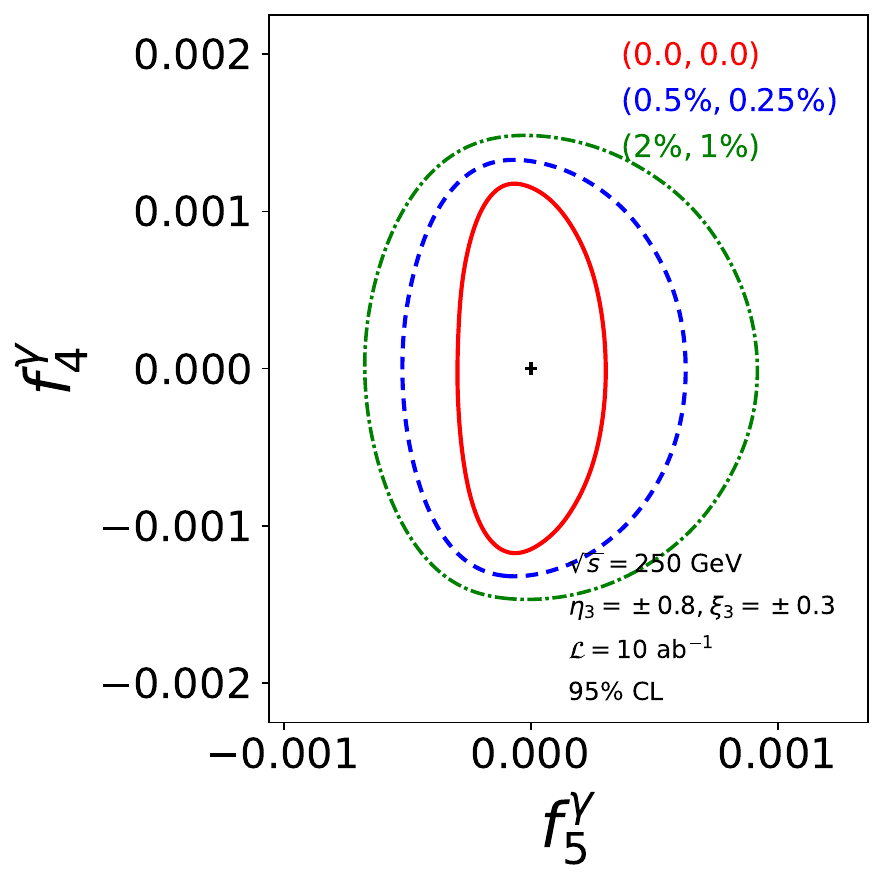}
    \includegraphics[width=0.32\textwidth]{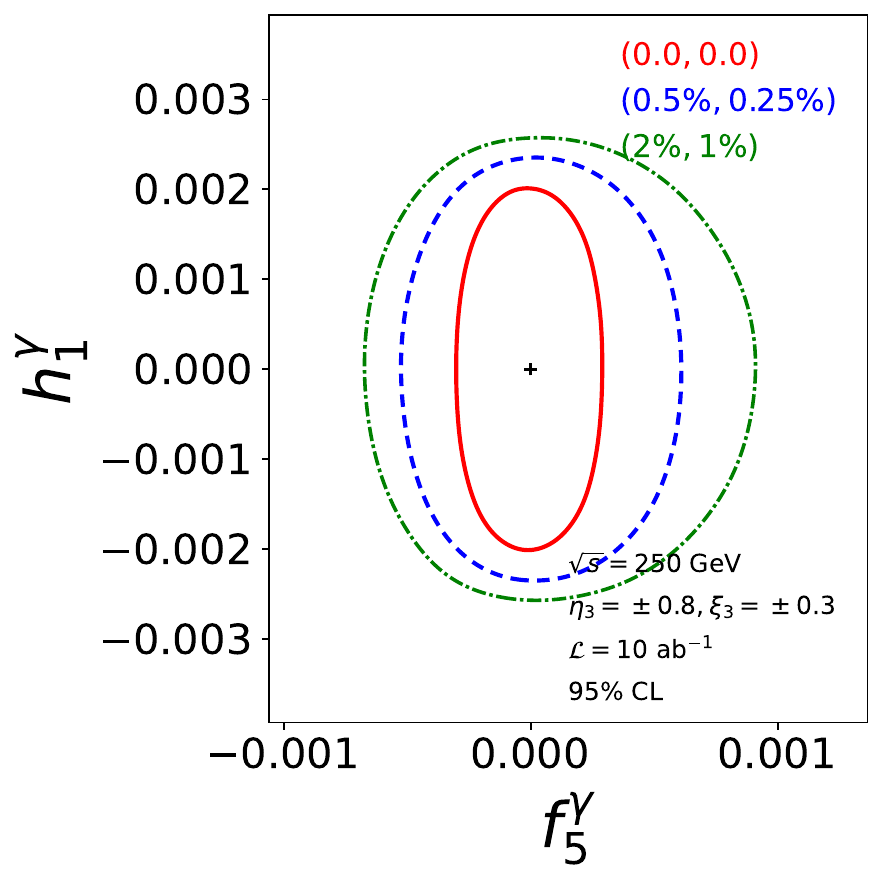}
    \includegraphics[width=0.32\textwidth]{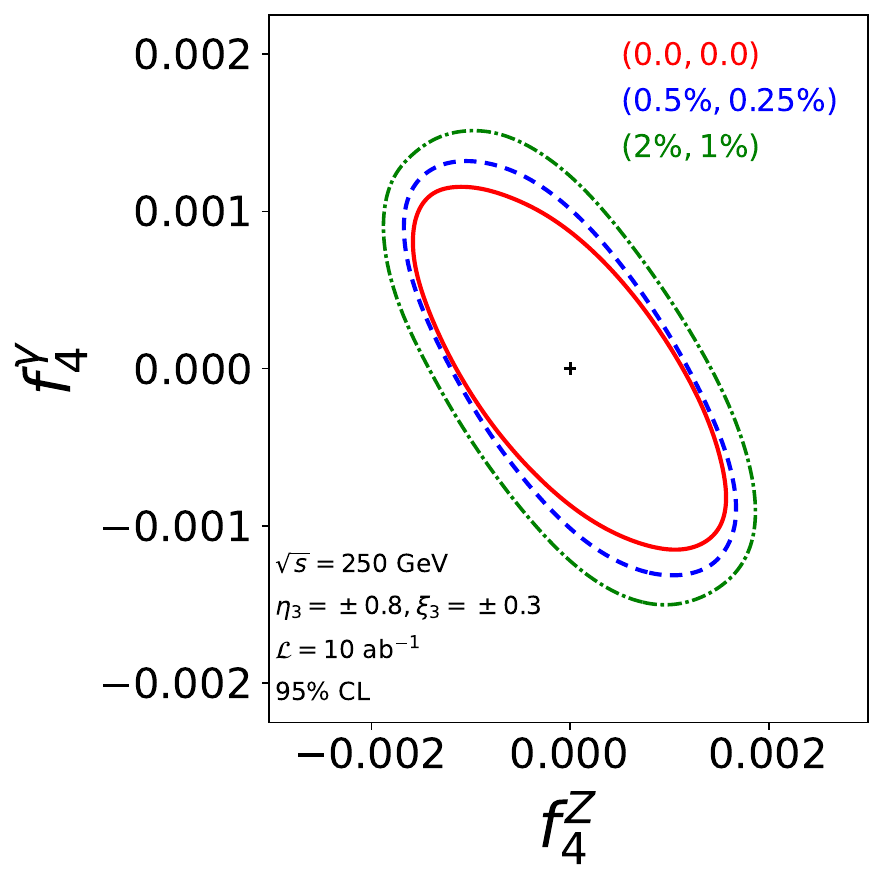}
    \includegraphics[width=0.32\textwidth]{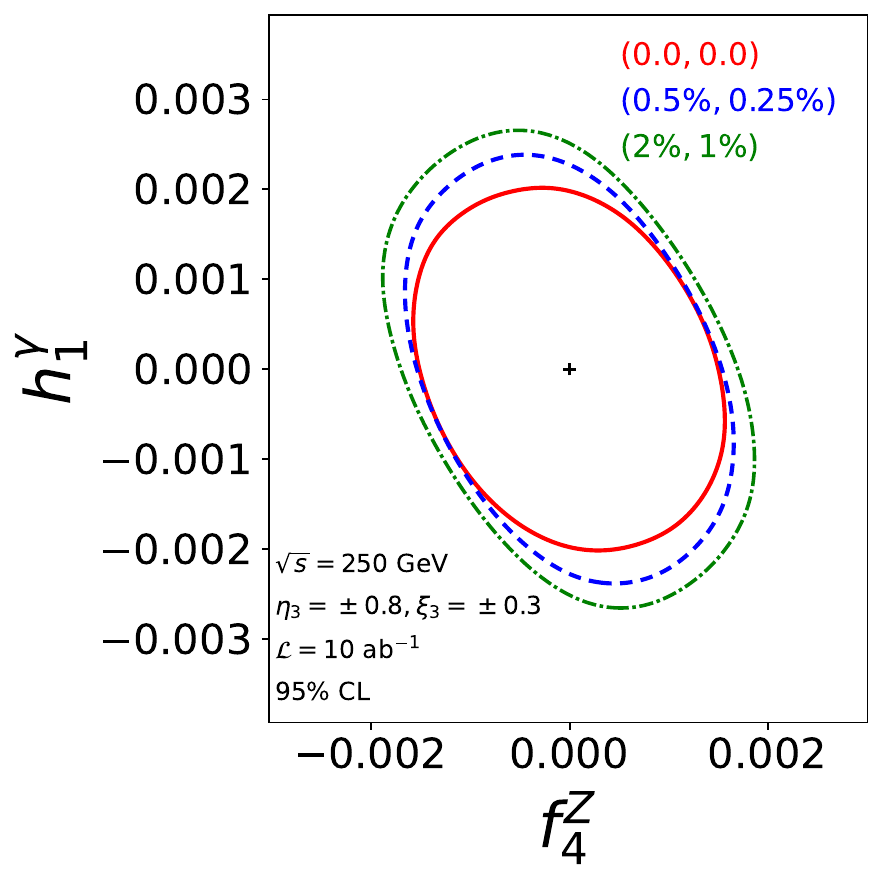}
    \includegraphics[width=0.32\textwidth]{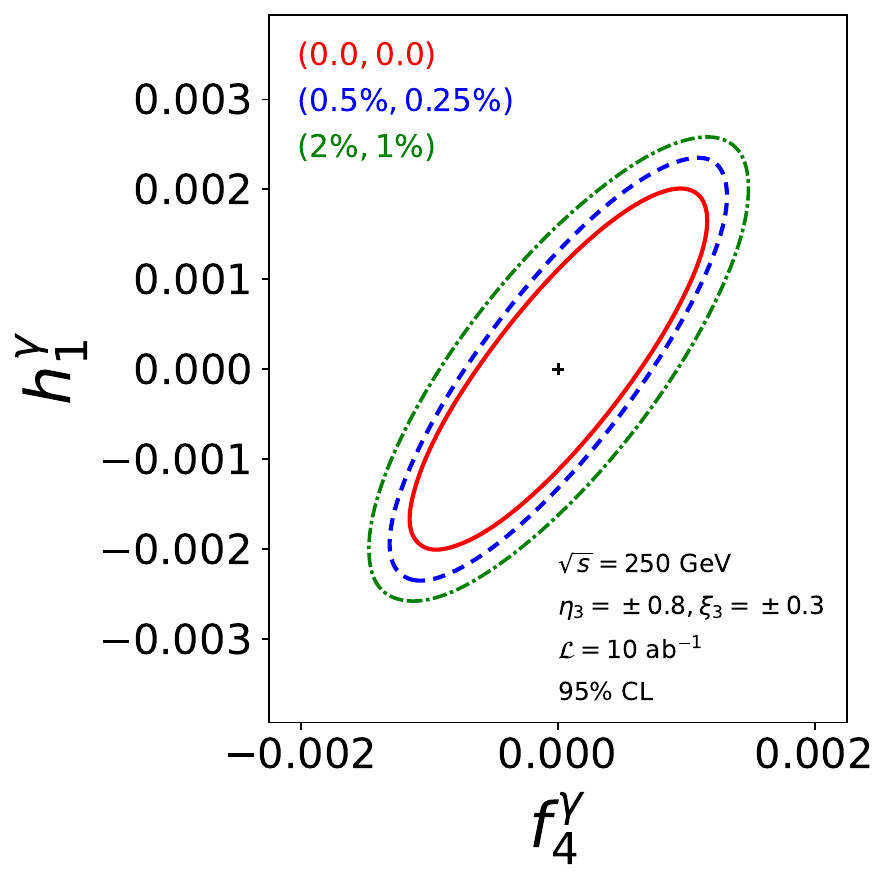}
   \caption{\label{fig:anom2dsyst}Two dimensional projections of marginalized anomalous couplings at $95\%$ CL for varying systematic error and integrated luminosity of 10 ab$^{-1}$.}
\end{figure*}
\section{Conclusion}
\label{sec:conclude}
In conclusion, we probe two processes viz. $e^-e^+\to ZZ\to4l$ and $e^-e^+\to Z\gamma\to 2l\gamma$ for upcoming International Linear Collider~(ILC) at center of mass energy of $250$~GeV with polarized beams in presence of four dimension-8 operators given in Eq.~(\ref{eqn:dim8op}). These operators generates anomalous neutral triple gauge couplings~($Z^\star ZZ,\gamma^\star ZZ,\gamma^\star Z\gamma$). To constrain the relevant Wilson coefficient, we construct polarization asymmetries of $Z$ boson in case of $Z\gamma$ and for $ZZ$ additional spin correlation asymmetries were also constructed. These observables were further divided into eight bins of $\cos\theta_Z$, where $\theta_Z$ is the angle between $Z$ boson and the beam axis in the lab frame. In case of $Z\gamma$ process, there exist eight asymmetries in each bin where five are CP-even and three are CP-odd. While for $ZZ$ process there are 80 asymmetries in each bin out of which 44 are CP-even and 36 are CP-odd. These CP-odd observables can be used to measure CP-violation in the production process. We used all these asymmetries along with cross~section to put a simultaneous constrain on anomalous couplings using MCMC analysis. The final limits are obtained by combining $Z\gamma$ and $ZZ$ process with two set of beam polarization at the level of $\chi^2$.\\
The limits on anomalous couplings $c_i$ except for $c_{BW}$ are dominated by the $Z\gamma$ process. The one parameter limits obtained in this work are tighter in the case of $ZZ$ process, while for the $Z\gamma$ process, the limits are comparable to the LHC limits listed in Table~\ref{tab:const}. The anomalous couplings contain momentum-dependent terms normalized by the electroweak scale, making them more prominent at higher energy. Analysis at LHC also enjoys higher cross~section for both $ZZ$ and $Z\gamma$ process; thus, variable like cross~section alone provides significant sensitivity to anomalous couplings. Thus, using many observables like polarizations and spin correlations becomes necessary to compensate for the low energy and cross~section at ILC. Those observables could probe the CP structure of new physics. The current analysis could be extended by inculcating semi-leptonic and hadronic decay channels of $Z$ bosons.


\begin{acknowledgments}
A. Subba acknowledge University Grant Commission, Govt. of India, New Delhi for financial support through UGC-NET Senior Research Fellowship.
\end{acknowledgments}
\appendix
\section{Anomalous couplings $f_i^V, h_i^V$}
\label{sec:app1}
We use Eq.~(\ref{eqn:EFTLag1} - \ref{eqn:EFTLag4}) to obtain the marginalized limits on $f_i^V$, and $h_i^V$ by MCMC global fits. We list the marginalized limits on four independent anomalous couplings in Table \ref{tab:95bcilag} for different values of luminosity and systematic errors. In order to understand the behaviour of the anomalous couplings in presence of varying systematic errors, we show the marginalized two dimensional projection of four independent anomalous couplings at $95\%$ CL in Fig.~\ref{fig:anom2dsyst}. The figures are obtained for different set of systematic errors and integrated luminosity of 10~fb$^{-1}$. The confidence interval for $f_5^\gamma$ improved by a factor of $\approx$ 2.68 on decreasing the systematic from ($2.0\%,1.0\%$) to zero. While for the $CP$-odd couplings, the confidence interval on $f_4^Z$ gets tighter by a factor of $\approx$ 1.7 and similarly for $f_4^\gamma,$ and $g_1^\gamma$ becomes better by a factor of $\approx$ 1.3. The number suggest that the at the value of luminosity of 10~fb$^{-1}$, the limits become blind to the systematic errors except in case of one $CP$-even coupling. Similarly, if the luminosity increases by a factor of 100 while keeping the conservative value of systematic error, the limits on all the anomalous couplings gets better by a factor of approximately 3. The limits on these couplings can be probed to better precision with higher center of mass energy and increasing the polarization degree of positron which is one of the technological challenge for future electron-positron Collider.
\begin{table*}
    \caption{\label{tab:95bcilag}The translated marginalized $95\%$ CL of anomalous couplings~$f_i^V,h_i^V$~($10^{-3}$) at $\sqrt{s} = 250$ GeV with beam polarization~($\eta_3,\xi_3$) = ($\pm0.8,\pm0.3$) from MCMC global fits for different values of luminosities and systematic error. }
    \centering
    \renewcommand{\arraystretch}{1.4}
    \begin{tabular}{llllll}\\\hline\hline    
         $\mathcal{L}$~(fb$^{-1}$)&\quad\quad$(\epsilon_\sigma,\epsilon_A)$&\quad\quad$f_5^\gamma$& \quad\quad $f_4^Z$ &\quad\quad $f_4^\gamma$ &\quad\quad $h_1^\gamma$ \\ \hline         
         &$(0.0,0.0)$&\quad\quad $[-1.43,+1.52]$ & \quad\quad$[-5.82,+5.50]$ &\quad\quad$[-3.68,+3.66]$&\quad\quad$[-6.34,+6.26]$\\
         100&($0.5\%,0.25\%$)&\quad\quad$[-1.45,+1.55]$&\quad\quad$[-5.60,+5.52]$&\quad\quad$[-3.68,+3.66]$&\quad\quad$[-6.35,+6.27]$\\
         &$(2.0\%,1.0\%)$&\quad\quad$[-1.57,+1.72]$&\quad\quad$[-5.69,+5.65]$&\quad\quad$[-3.71,+3.67]$&\quad\quad$[-6.43,+6.32]$\\ \hline
         
         &$(0.0,0.0)$&\quad\quad$[-1.00,+1.05]$& \quad\quad$[-4.10,+4.06]$&\quad\quad$[-2.77,+2.77]$&\quad\quad$[-4.72,+4.71]$\\
         $300$&$(0.5\%,0.25\%)$&\quad\quad$[-1.05,+1.12]$&\quad\quad$[-4.12,+4.10]$&\quad\quad$[-2.79,+2.79]$&\quad\quad$[-4.77,+4.75]$\\
         &$(2.0\%,1.0\%)$&\quad\quad$[-1.22,+1.38]$&\quad\quad$[-4.26,+4.24]$&\quad\quad$[-2.81,+2.81]$&\quad\quad$[-4.87,+4.84]$\\ \hline
         
         &$(0.0,0.0)$&\quad\quad$[-0.65,+0.66]$&\quad\quad$[-2.87,+2.86]$&\quad\quad$[-1.99,+2.01]$&\quad\quad$[-3.38,+3.38]$\\
         1000&$(0.5\%,0.25\%)$&\quad\quad$[-0.74,+0.80]$&\quad\quad$[-2.91,+2.90]$&\quad\quad$[-2.02,+2.03]$&\quad\quad$[-3.47,+3.46]$\\
         &$(2.0\%,1.0\%)$&\quad\quad$[-0.92,+1.12]$&\quad\quad$[-3.05,+3.04]$&\quad\quad$[-2.07,+2.08]$&\quad\quad$[-3.60,+3.60]$\\ \hline
         
         &$(0.0,0.0)$&\quad\quad$[-0.42,+0.41]$&\quad\quad$[-2.02,+2.00]$&\quad\quad$[-1.44,+1.45]$&\quad\quad$[-2.44,+2.44]$\\
         3000&$(0.5\%,0.25\%)$&\quad\quad$[-0.56,+0.63]$&\quad\quad$[-2.07,+2.06]$&\quad\quad$[-1.50,+1.51]$&\quad\quad$[-2.61,+2.61]$\\
         &$(2.0\%,1.0\%)$&\quad\quad$[-0.74,+0.08]$&\quad\quad$[-2.18,+2.18]$&\quad\quad$[-1.58,+1.59]$&\quad\quad$[-2.76,+2.76]$\\ \hline
         
         &$(0.0,0.0)$&\quad\quad$[-0.25,+0.24]$&\quad\quad$[-1.29,+1.27]$&\quad\quad$[-0.94,+0.95]$&\quad\quad$[-1.64,+1.65]$\\
         10000&$(0.5\%,0.25\%)$&\quad\quad$[-0.44,+0.50]$&\quad\quad$[-1.35,+1.35]$&\quad\quad$[-1.08,+1.08]$&\quad\quad$[-1.94,+1,94]$\\
         &$(2.0\%,1.0\%)$&\quad\quad$[-0.57,+0.74]$&\quad\quad$[-1.51,+1.50]$&\quad\quad$[-1.22,+1.23]$&\quad\quad$[-2.14,+2.14]$\\
                 \hline\hline
    \end{tabular}    
    \renewcommand{\arraystretch}{1.4}
\end{table*}

\bibliography{refer}
\end{document}